\documentclass[preprint,authoryear,12pt]{elsarticle}
\usepackage{epsfig}
\usepackage{graphicx}
\def\figfigincl#1#2#3{\includegraphics[width=#1]{figures/#2.eps}%
    \caption{\small #3}\label{fig:#2}}

\journal{Icarus}
\begin{document}

\title{On the Ages of Resonant, Eroded and Fossil Asteroid Families}
\author[pi]{Andrea Milani}
\ead{milani@dm.unipi.it}
\author[ac]{Zoran Kne\v zevi\'c}
\author[oca]{Federica Spoto} 
\author[to]{Alberto Cellino}
\author[bu]{Bojan Novakovi\'c}
\author[bg]{Georgios Tsirvoulis}
\address[pi]{Dipartimento di Matematica, Universit\`a di Pisa,
        Largo Pontecorvo 5,56127 Pisa, Italy}
\address[ac]{Serbiam Academy of Sciences and Arts, Knez Mihajlova 35, 11000 Belgrade, Serbia}
\address[bg]{Astronomical Observatory, Volgina 7,
         11060 Belgrade 38, Serbia}
\address[oca]{Laboratoire Lagrange, Universit\`e C\^ote d'Azur,
  Observatoire de la C\^ote d'Azur, CNRS}
\address[to]{INAF--Osservatorio Astrofisico di Torino,
10025 Pino Torinese, Italy}
\address[bu]{Department of Astronomy, Faculty of Mathematics,
  University of Belgrade, Studentski trg 16, 11000 Belgrade, Serbia}
\begin{abstract}
In this work we have estimated 10 collisional ages of 9 families for
which for different reasons our previous attempts failed. In general,
these are difficult cases that required dedicated effort, such as a
new family classifications for asteroids in mean motion resonances, in
particular the $1/1$ and $2/1$ with Jupiter, as well as a revision of
the classification inside the $3/2$ resonance. 

Of the families locked in mean motion resonances, by employing a
numerical calibration to estimate the Yarkovsky effect in proper
eccentricity, we succeeded in determining ages of the families of
(1911) Schubart and of the ``super-Hilda'' family, assuming this is
actually a severely eroded original family of (153) Hilda.  In the
Trojan region we found families with almost no Yarkovsky evolution,
for which we could compute only physically implausible ages. Hence, we
interpreted their modest dispersions of proper eccentricities and
inclinations as implying that the Trojan asteroid families are fossil
families, frozen at their proper elements determined by the original
ejection velocity field.  We have found a new family, among the
Griquas locked in the 2/1 resonance with Jupiter, the family of
(11097) 1994 UD1.

We have estimated the ages of 6 families affected by secular
resonances: families of (5) Astraea, (25) Phocaea, (283) Emma, (363)
Padua, (686) Gersuind, and (945) Barcelona. By using in all these
cases a numerical calibration method, we have shown that the secular
resonances do not affect significanly the secular change of proper a.

For the family of (145) Adeona we could estimate the age only after
removal of a number of assumed interlopers.

With the present paper we have concluded the series dedicated to the
determination of asteroid ages with a uniform method. We computed the
age(s) for a total of 57 families with $>100$ members. For the future
work there remain families too small at present to provide reliable
estimates, as well as some complex families (221, 135, 298) which may
have more ages than we could currently estimate. Future improvement of
some already determined family ages is also possible by increasing
family membership, revising the calibrations, and using more reliable
physical data.

\end{abstract}

\maketitle


\newpage

\section{Introduction}
\label{s:intro}

In our previous work (\cite{bigdata}, hereinafter referred to as Paper
I, and \cite{namur_update}, Paper II), we have introduced new methods
to classify asteroids into families, applicable to an extremely large
dataset of proper elements, to update continuously this
classification, and to estimate the collisional ages of large
families. In later work (\cite{ages}, Paper III, and in
\cite{iaus318}, Paper IV), we have systematically applied a uniform
method (an improvement of that proposed in Paper I) to estimate
asteroid family collisional ages, and solved a number of problems of
collisional models, including cases of complex relationship between
dynamical families (identified by clustering in the proper elements
space) and collisional families (formed at a single time of
collision).

In this paper we solve several difficult cases of families for which
either a collisional model had not been obtained, e.g., because it was
not clear how many separate collisions were needed to form a given
dynamical family, or because our method based on V-shapes (in the
plane with coordinates proper semimajor axis $a$ and inverse of
diameter $1/D$) did not appear to work properly. In many cases the
difficulty had to do with complex dynamics, such as the effects of
resonances (either in mean motion or secular). In other cases the
problem was due to the presence of interlopers (members of the family
by the automatic classification but not originated from the same parent
body) which can be identified by using physical data, including
albedo, color indexes and absolute magnitudes.

Thus, to assess the level of success of the research presented in this
paper, the reader should take into account that all the families
discussed in this paper have been selected by the failure of our
previous attempts to find a reasonable collisional model and/or to
estimate an age. In this paper we discuss new family classifications
for asteroids in mean motion resonances, in particular the $1/1$ and
$2/1$ with Jupiter, plus a revision and critical discussion of the
classification inside the $3/2$ resonance. Overall we have estimated
$10$ additional ages for $9$ families. Almost all the succesful
computations have required some additional effort, on top of using the
methods developed in our previous papers. Two examples: first, removal
of interlopers has played a critical role in several cases, to the
point that $2$ of the $9$ families used for age estimation have seen
removal of the namesake asteroid as an interloper, resulting in change
of the family name\footnote{We are using the traditional asteroid
  family naming convention, by which the family is named after the
  lowest numbered member: if the lowest numbered is an interloper, the
  second lowest numbered becomes the namesake.}; second, the
interaction of the Yarkovsky effect with resonances has very different
outcomes: outside resonances, the semimajor axis undergoes a secular
drift, while inside a mean motion resonance the semimajor axis is
locked and other elements can undergo a secular drift. Inside a
secular resonance the drift in semimajor axis appears not to be
significantly affected, as shown by our numerical tests.

The paper is organized as follows: in Section~\ref{s:meanmot} we deal
with the specific problems of families formed by asteroids locked
inside the strongest mean motion resonances with Jupiter, namely $3/2$
(Hildas), $1/1$ (Trojans), and $2/1$ (Griquas). In
Section~\ref{s:secres} we discuss the families affected by secular
resonances (involving the perihelia and nodes of the asteroid and the
planets Jupiter and/or Saturn).
In Section~\ref{s:complex} we present two successful interpretations
of families with strange shapes and many interlopers. In
Section~\ref{s:obstest} we discuss the families for which we either
have not found a consistent collisional model, or we have found a
model (and computed an age) but there are still problems requiring
dedicated observational efforts.
In Section~\ref{s:conclusions} we draw some conclusion, not just from
this paper but from the entire series, in particular discussing the
limitations to the possibility of further investigating the
collisional history of the asteroid belt with the current data set.

The numerical data in connection with the computation of ages
presented in this paper are collected in the Tables~1--8, analogous to
those used in Papers III and IV: all the tables are given in the
Appendix. Tables~\ref{tab:tablefite} and \ref{tab:tablefita} describe
the fit region in the $(e, 1/D)$ and $(a, 1/D)$ planes, respectively,
where $D$ is the diameter in km.  Table~\ref{tab:tablealbedo} contains
the albedo data from various sources, used to estimate $D$.
Tables~\ref{tab:tableslopese} and \ref{tab:tableslopesa} contain the
results of the fit for slopes of the V-shapes, again in the two
planes. Table~\ref{tab:tableyarkoparam} contains the data to compute
the Yarkovsky calibration (see Papers I and III), and finally
Tables~\ref{tab:tableages_dedt} and \ref{tab:tableages_dadt} contain
the calibrations, the estimated ages and their uncertainties.
Tables~\ref{tab:tablefita}, \ref{tab:tablealbedo},
\ref{tab:tableslopesa}, \ref{tab:tableyarkoparam} and
\ref{tab:tableages_dadt} are partitioned by horizontal lines into
sections for families of type fragmentation, of type cratering, and
with one side only. We have found no additional young ages ($< 100$
My).

For the sake of brevity, unless absolutely necessary, we do
not specifically quote in the text these tables and data they contain,
but the reader is encouraged to consult them whenever either some
intermediate result is needed, or some quantitative support of the proposed
explanation is required.

\section{Families in mean motion resonances with Jupiter}
\label{s:meanmot}

Mean motion resonances with Jupiter give rise to the Kirkwood gaps,
where the asteroid main belt has essentially empty regions
corresponding to the most chaotic orbits: the main gaps correspond to
the resonances $3/1$ and $5/2$. For the order 1 resonances $3/2$ and
$2/1$ there are gaps created in the regions of strongly chaotic orbits
but also islands of relative stability, where families can be
found. The $1/1$ resonance contains a large stable region containing
the two Trojan swarms (preceding and following Jupiter).

\subsection{3/2 resonance: the Hilda region}
\label{s:hilda}

\begin{figure}[h!]
\figfigincl{12 cm}{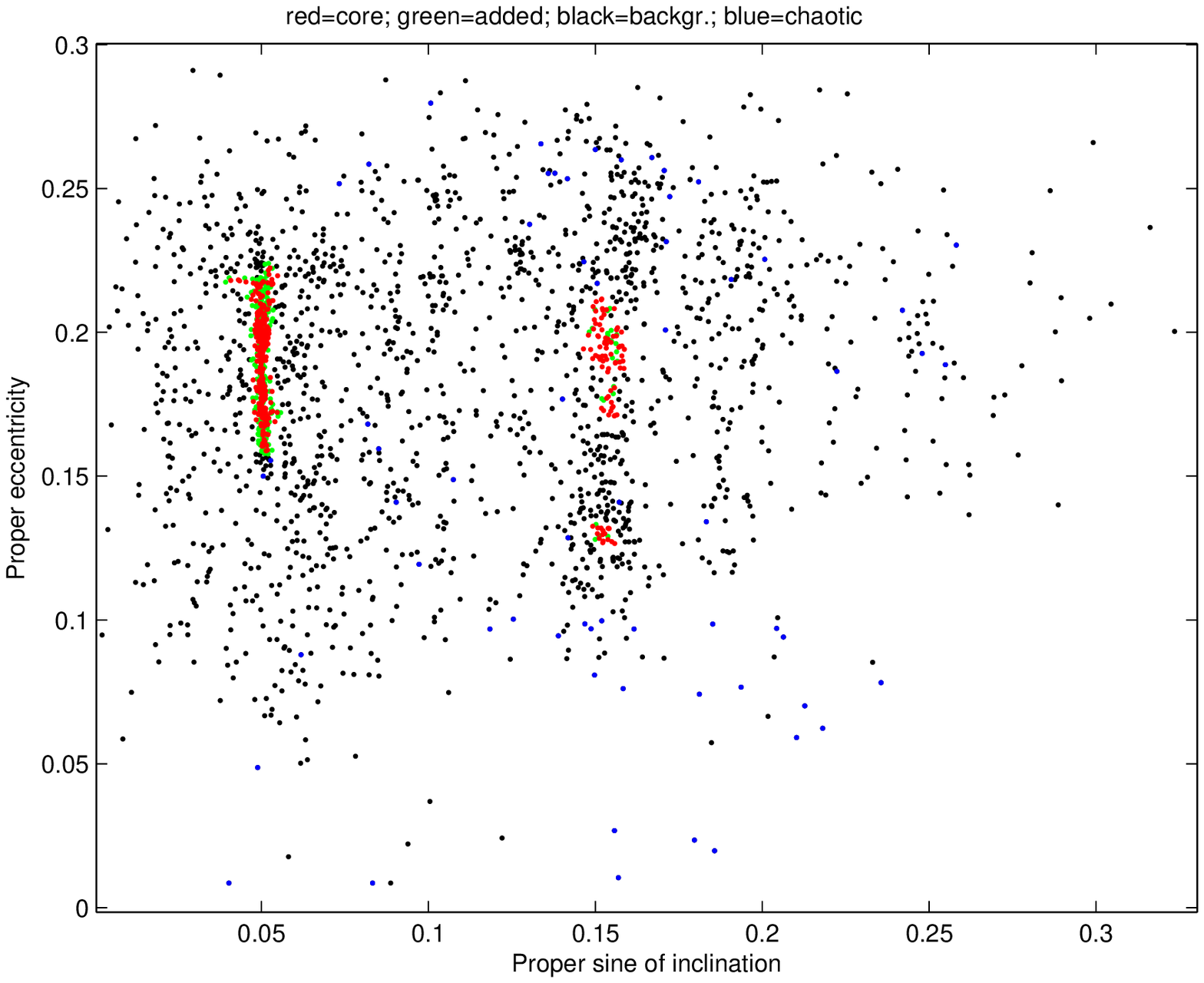}{The Hilda region projected on the
  proper $(\sin{I}, e)$ plane. The visible concentrations are the
  family of (1911) Schubart on the left (around proper $\sin{I}=0.05$)
  and the eroded family of (153) Hilda near the center (around
 proper $\sin{I}=0.15$). The different number density indicates a different
  collisional and Yarkovsky history. Red points mark family members
  obtained by the standard HCM procedure, while green are members
  attributed in later updates (Papers I and II), the blue are
  chaotic.}
\end{figure}

The $3/2$ mean motion resonance with Jupiter contains a large region
in which the critical argument $2\lambda-3\lambda_J+\varpi$ can
librate for a very long time. The computation of proper elements
specifically for the $3/2$ resonant orbits is possible
\citep{schubart}, but not really necessary for the purpose of family
classification. The reason is that most information is contained in
the proper elements $(e,\sin{I})$ (see Figure~\ref{fig:hilda_region}),
while the value of the smoothed and averaged $a$ is sharply
constrained close to the libration center (see
Figure~\ref{fig:hildareg_ae}). Thus we are using in this region
``semi-proper'' elements $(a, e, \sin{I})$ computed with the same
algorithm used for the non-resonant case, and some families 
can be identified by the HCM method all the same.

Two clusters are visible by eye in Figure~\ref{fig:hilda_region}, at
low proper $\sin{I}$ the very well defined family of (1911) Schubart
and at higher proper $\sin{I}$ a clan, that is a number of small families
which may be interpreted as the remains of an eroded, very ancient
family, which could have included the largest asteroid in the region,
(153) Hilda.

\begin{figure}[t]
\figfigincl{12 cm}{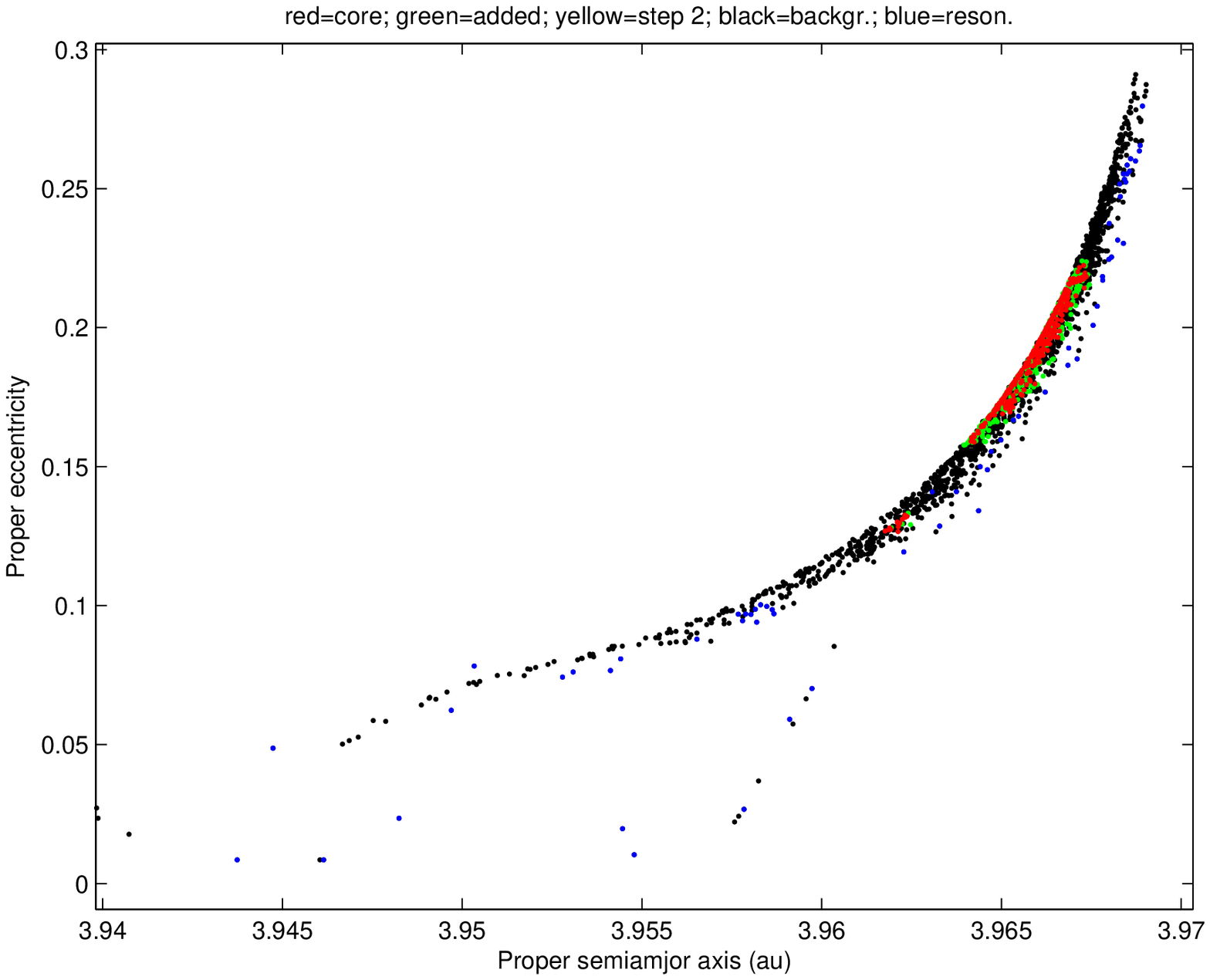}{The Hilda region projected on the
  proper $(a, e)$ plane. The averaging used to compute synthetic
  proper elements produces a narrow strip around the libration center
  line, the Yarkovsky driven secular changes can take place only along
  this line. Due to the blown-up scale in proper semimajor axis the
  libration center line appears strongly curved, but in reality the
  effect is small.}
\end{figure}

\subsubsection*{The Schubart family}
\label{s:schubart}

The most prominent family we have found in the Hilda region with our
methods (Paper I, II) is the family of (1911) Schubart now grown to
508 members (of which 136 multiopposition, added in Paper
IV).\footnote{This family had already been reported in
  \cite{schubart}.} The methods of Paper III to compute the age are
not strictly applicable to this case, because the range in $a$ is very
narrow, only $0.035$ au. Indeed a V-shape in the plane with
coordinates proper $a$ and $1/D$ is well defined but does not describe
the amount of change in the proper elements due to the Yarkovsky
effect: it is constrained by the slope of the libration curve in the
region where the family is.

The Hildas remain locked in the $3/2$ resonance (see
Figure~\ref{fig:hildareg_ae}), thus the Yarkovsky effect pushes the
eccentricity either up (for direct spin) or down (for retrograde spin)
\citep{bottke2002}. This results in a V-shape in the plane of proper $e$ and
$1/D$ (Figure~\ref{fig:1911_vshapee}), which can be used to estimate
the age, provided the Yarkovsky calibration for $de/dt$ is available.

\begin{figure}[t]
\figfigincl{14 cm}{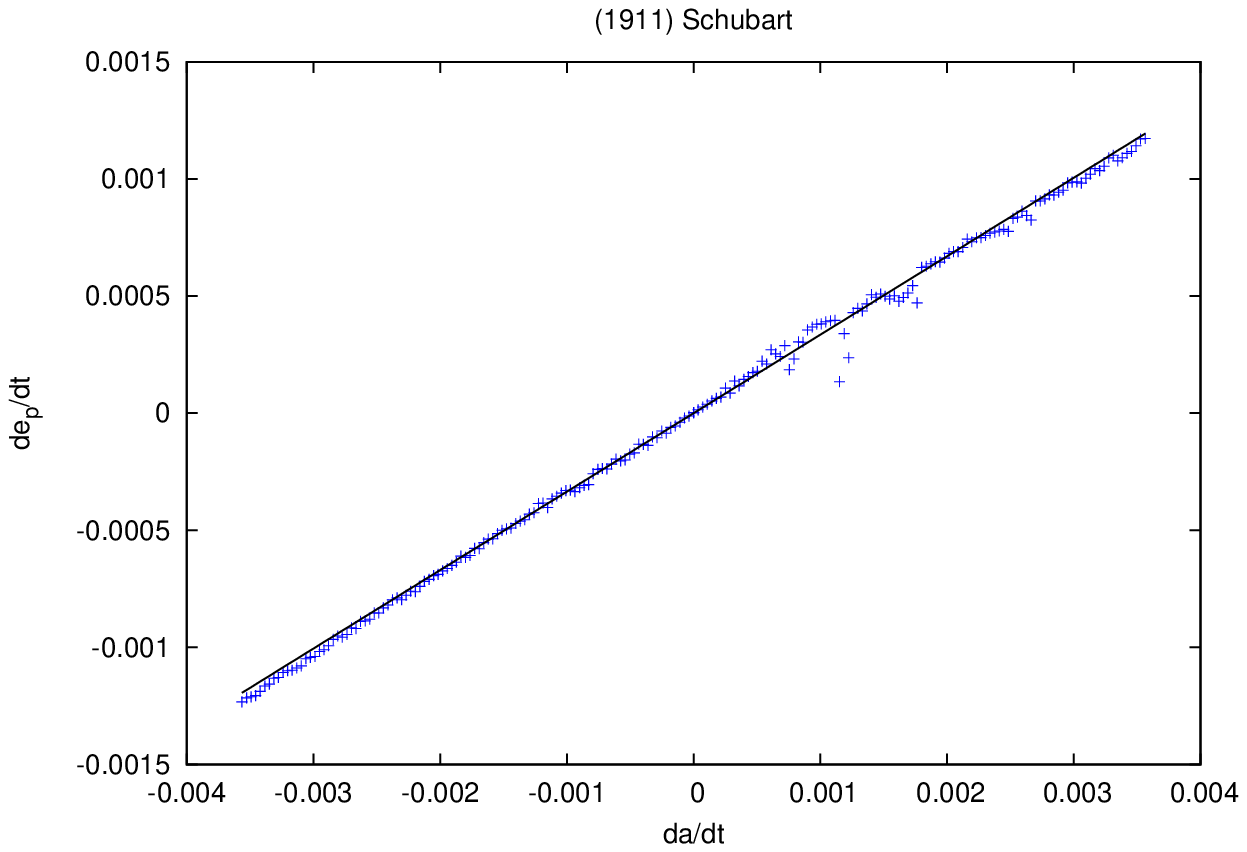}{The eccentricity calibration of the
  Yarkovsky effect inside the $3/2$ resonance, for the family of
  (1911) Schubart. On the horizontal axis, the value of $da/dt$ (in
  au/My) as it would have been outside of the resonance; on the
  vertical axis the value of $de/dt$ actually measured for a resonant
  orbit. The straight line is the regression line fitted to the $200$
  data points.}
\end{figure}

To obtain the calibration in $de/dt$ we have used numerical
experiments with a sample of $200$ clones of the orbit of (1911) with
values of non-resonant $da/dt(NR)$ assigned in the range $-3.6\times
10^{-9}<da/dt(NR)<3.6\times 10^{-9}$ au/y. The integrations were
performed for a time span of $10$ My, then proper elements were
computed for windows of $2$ My, shifting by $1$ My, giving a total of
$9$ data points for each clone. To these we fitted a $de/dt$ obtaining
the results shown in Figure~\ref{fig:schubart_calib}.  Then a
regression line was fitted in the $(da/dt(NR),de/dt)$ plane:
\[
\frac{de}{dt}=0.335 \cdot \frac{da}{dt}(NR)\ ;
\]
and the coefficient $0.335$ has to be used to convert
the $da/dt$ calibration, which we obtained with the same procedure
used in Paper I and III, into a $de/dt$ calibration.

\begin{figure}[h!]
\figfigincl{12 cm}{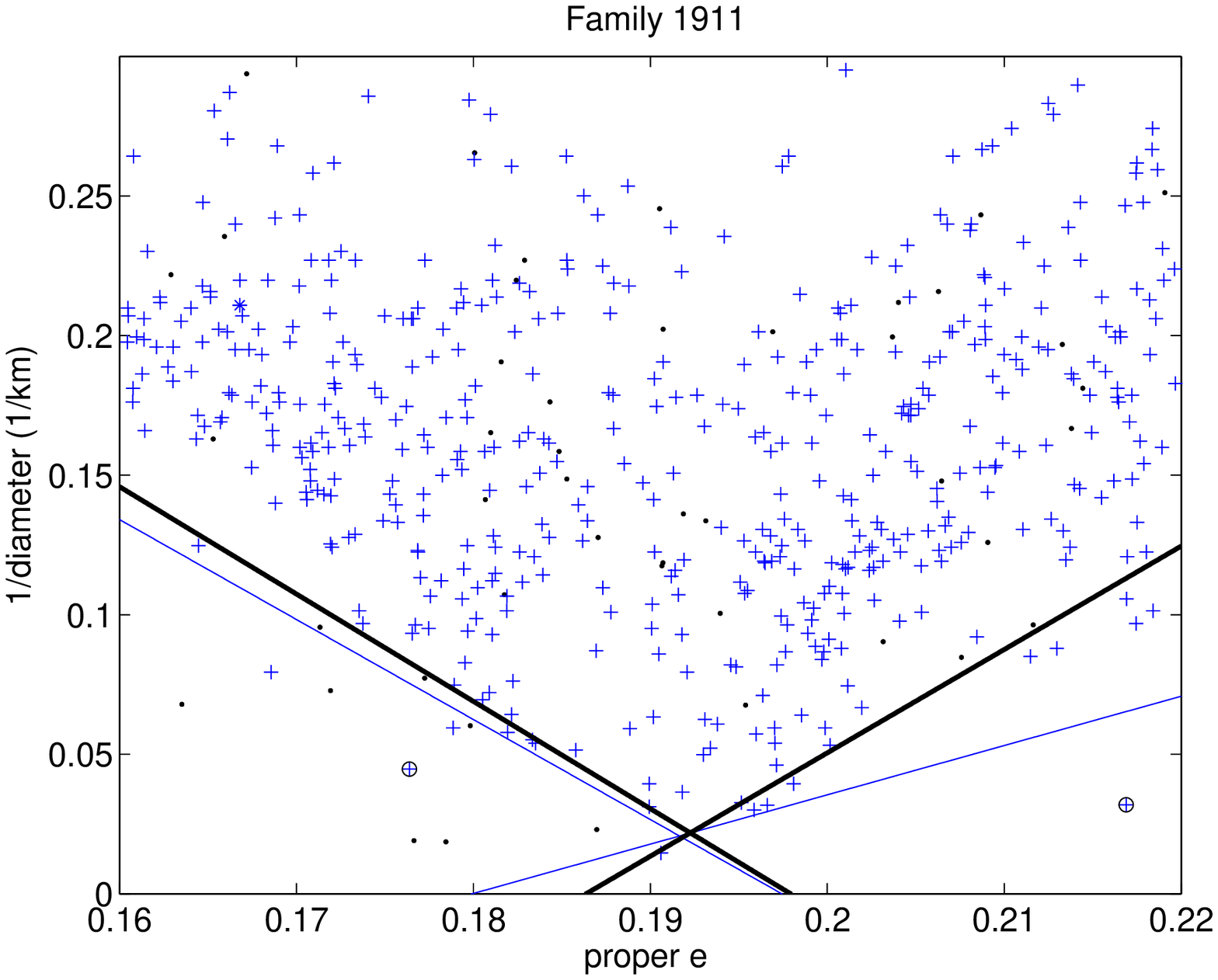}{The V-shape for the family of (1911)
  Schubart in the plane $(e, 1/D)$, with the fit lines (blue the first
  iteration, black the final one). The crosses are family members, the
  one with also a circle are outliers discarded from the V-shape fit;
  dots are non-members.}
\label{1911vshapee}
\end{figure}

The V-shape fit in ($e, 1/D$) plane (Figure~\ref{1911vshapee}) reveals
consistent slopes (see Table~\ref{tab:tableslopese})
and we can model this family as a single collisional family.

The Yarkovsky calibration converted in $e$ is $-1.68\times 10^{-10}$
in 1/y for the IN side and $1.72\times 10^{-10}$ for the OUT side,
hence the age estimate is $1,547\pm 492$ for IN, $1,566\pm 484$
for OUT side, in My. Note that the uncertainty is completely dominated
by the relative calibration uncertainty of $0.3$, thus more work on
calibration is needed.

The slight curvature of the libration center line of
Figure~\ref{fig:hildareg_ae} results in a slow motion also in proper
$a$, but the secular $da/dt$ is $0.016$ times what it would be outside
the resonance. Thus, in principle, the age could be computed by
using the V-shape in $(a, 1/D)$ and the calibration above, with
similar results but poorer accuracy with respect to the results in the
$(e, 1/D)$ plane.

\subsubsection*{The Hilda core and other small families}

The high $\sin{I}$ portion of the $3/2$ resonance shows some density
contrast, see Figure~\ref{fig:hilda_region}: the rectangle with
$0.075<e<0.29$ and $0.14<\sin{I}<0.17$ contains $484$ asteroids with
proper elements. 

Our HCM run, performed in a single step in this region (see Paper I), finds
there two small families: 6124 with $73$ members, and 3561 with only
$19$.

However, if we definine first core families among the subpopulation
with magnitude $H<14.75$, we find a core family of (153) Hilda with
only $14$ members. This family is not recognized as significant in the
single step HCM because the limit number for significance, in this
case, would be set at $15$ members. Since this candidate family was
important and only marginally discarded, we have decided to use the
two step procedure, at least to assess the situation. Then
the second step, consisting in the attribution of fainter
objects to the existing core families, gives a surprisingly small
growth, namely the membership of family 153 increases only to $18$
members. The conclusion is that a family of (153) Hilda can be
considered to exist, but has a peculiar size distribution, strongly
depleted in small size members; this is not due just to observational
selection, since in the $3/2$ resonance $52\%$ of the asteroids with
proper elements have $H>14.75$.

The other families of (1911) Schubart, of (6124) Mecklenburg, not far
from 153, and of (3561) Devine at lower $a$ and $e$, are not affected
at all, that is their membership is the same both with the one step
and the two step procedure. 


At levels of distance above the one which is prescribed by HCM,
families 153 and 6124 merge at $70$ m/s, 3561 merges at $120$ m/s,
thus forming a very large ``super--Hilda'' family (Paper I, Section
4.2).  An even bigger extension of the Hilda family was proposed by
\cite{brozetal2011}, with a limiting distance of $140$ m/s; they also
claim that it is a very ancient family, with an age estimated at about
$4$ Gy.  This is a legitimate method to propose a family, although not
a way to prove that it is statistically significant.

\begin{figure}[h!]
\figfigincl{12 cm}{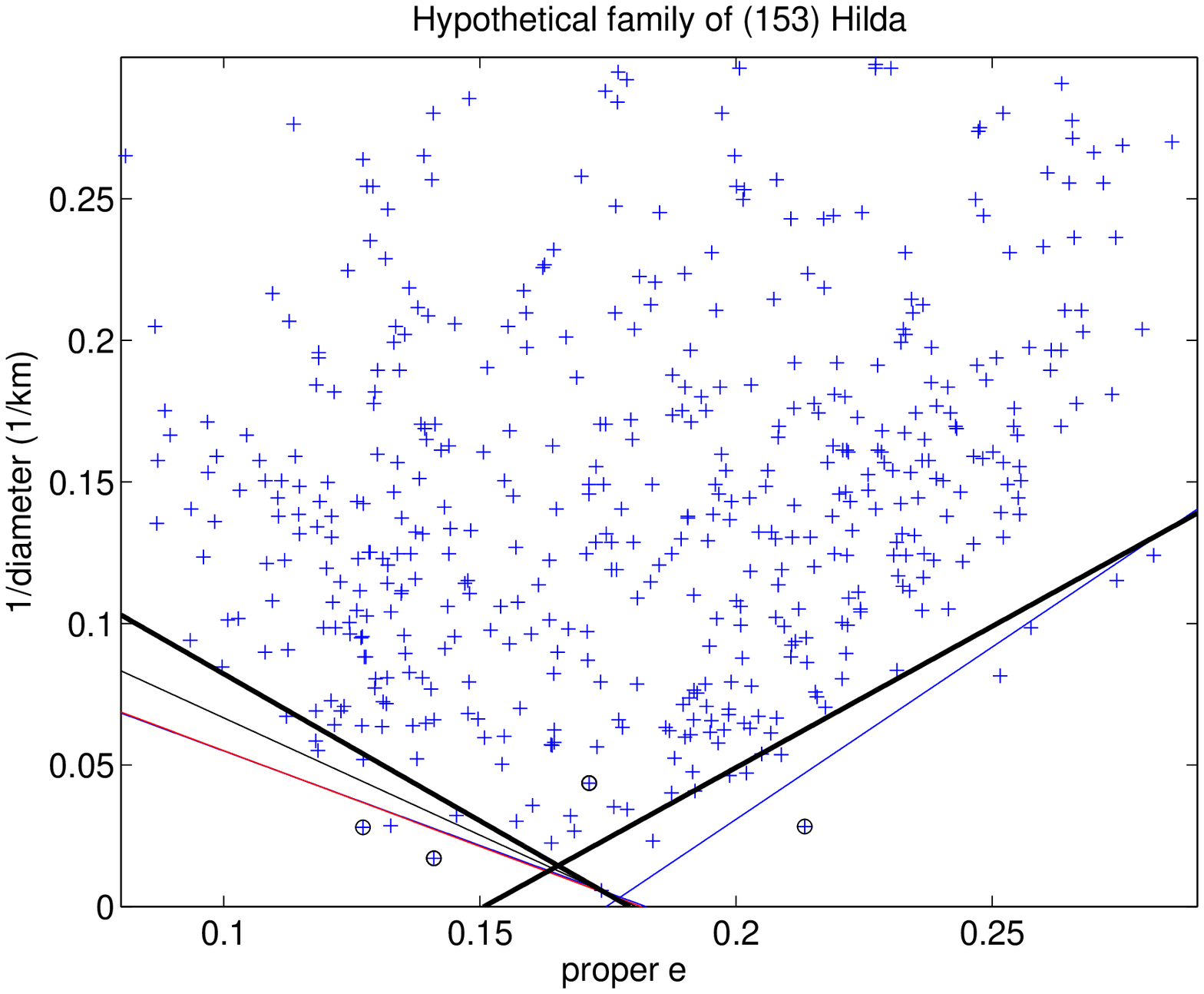}{The V-shape for the super-family of
  (153) Hilda in the plane $(e, 1/D)$.  Note that the V-shape is
  formed by comparatively large members, for the most part with $D>10$
  km.}
\end{figure}

Thus we have decided to investigate the properties of the super-Hilda
family, even if its significance cannot be proven by HCM.

The fact which is known about very ancient families is that they tend
to erode: secular drifts due to Yarkovsky, chaotic escape routes,
and collisional comminution must necessarily occur and affect all
members, but in a size-dependent way. The non-gravitational
perturbations typically are proportional to $1/D$, most resonances
containing escape routes become accessible as a result of Yarkovsky
secular drift, collisional lifetimes are shorter for smaller objects.
Anyway, whatever the cause, ancient families dissipate by
preferentially loosing the smaller members. 

Thus it is possible that a super-Hilda family did exist, but is now so
much depleted in small members that it cannot appear as a connected
density contrast (which is statistically significant), but gets
disconnected in three smaller islands.  The slope of the size
distribution of the $484$ asteroids in the box cited above is
$N(D)\sim 1/D^{-1.75}$, that is, extremely shallow.

If such a family existed, then it could still show a V-shape, in this
case of $3/2$ resonance in the proper $(e,1/D)$ plane.  Indeed,
Figure~\ref{fig:hilda_vshapee} does show a pronounced V-shape which
can be used to compute an age. In the same time, neither of the small
isolated groups does show any meaningful V-shape. The calibration of
$de/dt$ is obtained with exactly the same procedure as for family
1911, and with the similar result:
\[
\frac{de}{dt}=0.370 \cdot \frac{da}{dt}(NR)\ .
\]
Note that for the drift in $a$ due to the slope of the libration center
the coefficient (w.r. to the drift which would occur outside the
resonance) is also similar to the one for the family 1911: $0.020$.

The slopes (see Table~\ref{tab:tableslopese}) are consistent. The
Yarkovsky calibration converted in $e$ is $-1.83\times 10^{-10}$ in
1/y for the IN side and $1.99\times 10^{-10}$ for the OUT side. The
age estimate is $5,265\pm 1,645$ for the IN side, $5,039\pm 1,682$ for
OUT, in My; the uncertainty is again dominated by the $0.3$ relative
uncertainty in calibration.

Of course the nominal value and higher values in the formally
acceptable range are not possible, but the lowest part of the range of
uncertainty gives legitimate ages, older than any other we have
measured, and compatible with \cite{brozetal2011} results. Such an old
age estimate can indeed justify the very shallow size distribution;
however, with the current number of identified members we cannot claim
this is a strong confirmation of the age.

Note that, if this is really a very old family strongly depleted in small
members, we should have here a rare case in which we can expect that
future observations, extending the completeness limit to fainter objects,
will {\em not} produce a significant increase of the number of family
members, as opposite to the case of the large majority of asteroid families.
This shall be futher discussed in Section~\ref{s:superhilda}

\subsection{1/1 resonance: the Trojans}

\begin{figure}[h!]
\figfigincl{12 cm}{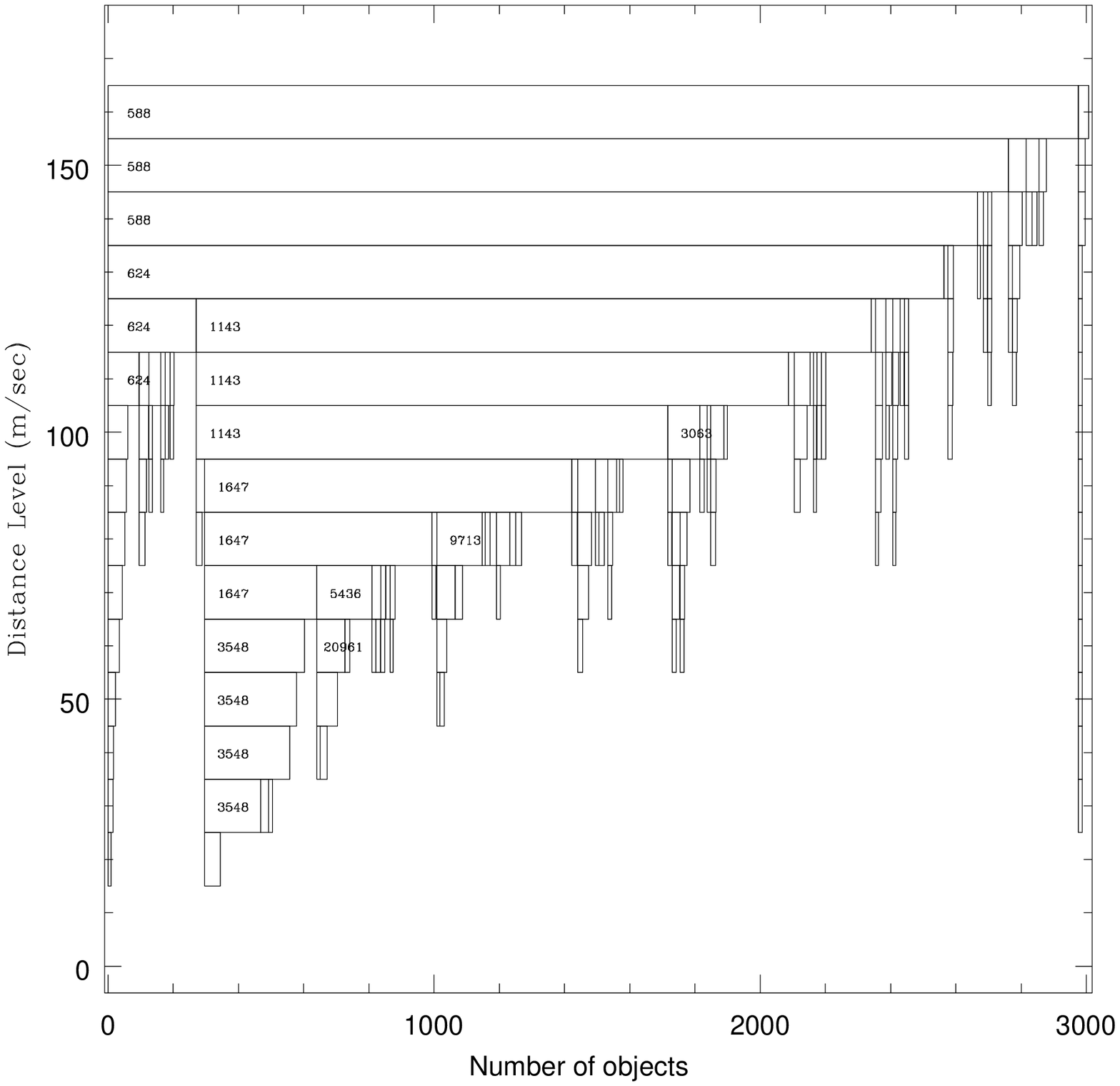}{Stalactite diagram for the 
Trojans of the L4 swarm. The value of $N_{lim}$ is $10$.
The distance level at which families are identified is $30$ m/s.}
\end{figure}
The Jupiter Trojans dataset includes now $3357$ objects surrounding L4
(the ``Greek camp''), and $1663$ surrounding L5 Lagrange point (the
``Trojan camp'').  The proper elements have been computed with the
synthetic method described in \cite{milani1992,milani1993}, and are
different in that the proper $a$ is replaced by the semi-amplitude
$\Delta a$ of the libration in semimajor axis due to the $1/1$
resonance\footnote{To display the two swarms separately, but on the
  same planes as the main belt, in our visualizer at {\tt
    http://hamilton.dm.unipi.it/astdys2/Plot/index.html}, we have used
  the values $a_J-\Delta a$ for L4, $a_J+\Delta a$ for L5.}; the
proper elements $e$ and $\sin I$ are computed in the same way as in
the main belt.

\begin{figure}[t!]
\figfigincl{12 cm}{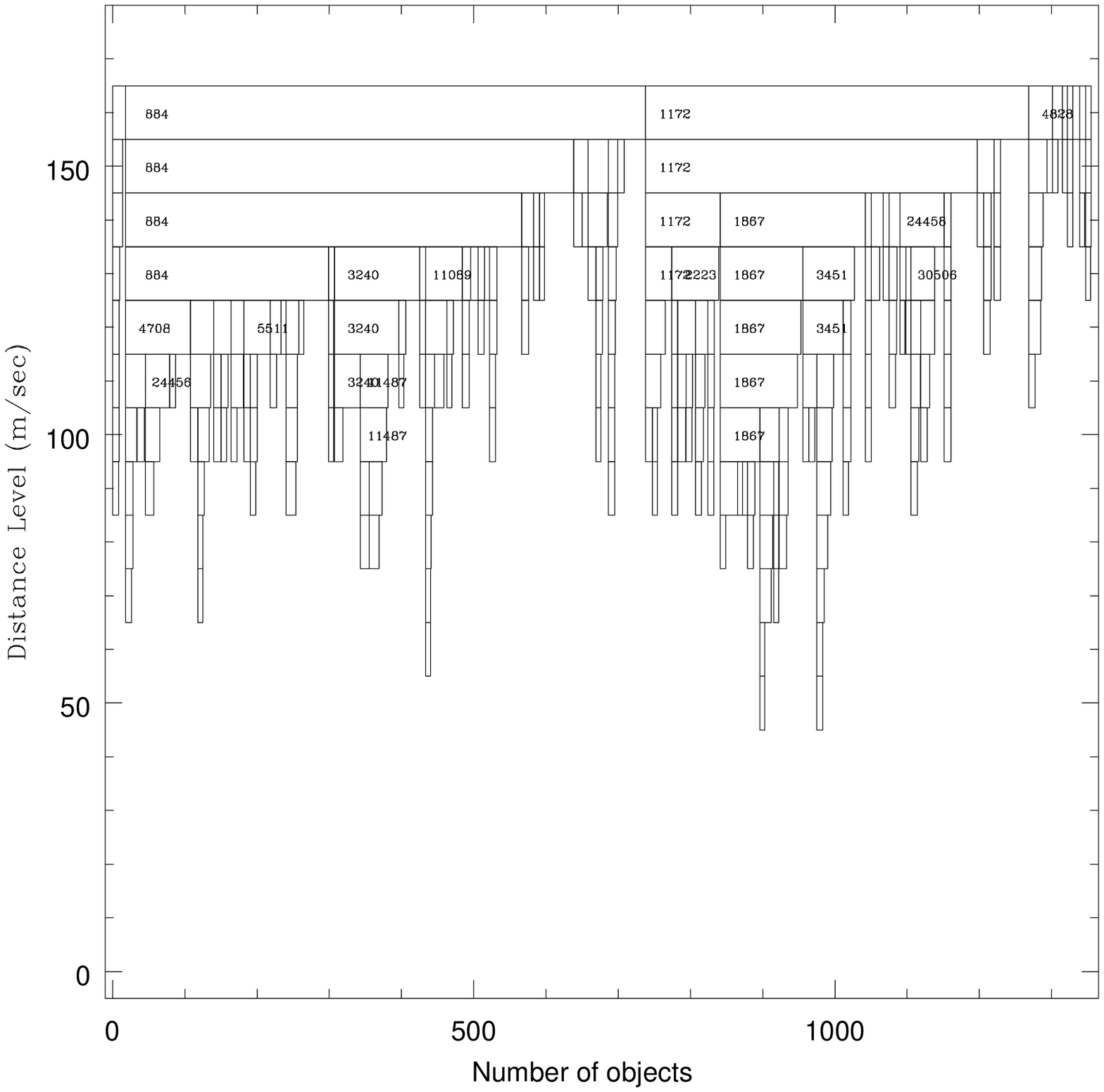}{Stalactite diagram for the 
Trojans of the L5 swarm. The value of $N_{lim}$ is $7$.
The distance level at which families are identified is $50$ m/s.}
\end{figure}

We have used, separately for the L4 and the L5 swarm, a single step
HCM method, with the same metric of \cite{zapetal90,zapetal95}, that
is using for the differences in $\Delta a$ the same $5/4$ coefficient used
for the differences in $a$ in the main belt (anyway, the differences
in $\Delta a$ play a minor role due to the fact that their range of
values is narrow: $0< \Delta a< 0.17$ au).

The stalactite diagram for the L4 objects, obtained by using a value
of $N_{lim} = 10$, is shown in Fig.~\ref{fig:stalatL4}. Since we
found for the Quasi Random Level a value of $40$ m/s, this means that
families must be found at $30$ m/s.  The stalactite diagram for the L5
Trojan asteroids is shown in Fig.~\ref{fig:stalatL5}.  The
adopted value of $N_{lim}$ for L5 is $7$ and the QRL is found at $60$
m/s.  L5 families are therefore defined at $50$ m/s.

By looking at the stalactite diagrams for L4 and L5, one sees a
multitude of thin and in several cases very deep stalactites emerging
from the bulk of the two populations.  For some of them, mainly in L4,
the unusual thing is just their thin and compact morphology.

This behaviour is not unexpected {\em a priori}. The dynamical
environment in the Jupiter Trojan swarms is such that the usual
mechanisms of dynamical dispersion of families at work in the main
belt, including primarily the Yarkovsky effect, are much weaker here,
even over long time scales.

As a consequence, swarms of fragments produced by collisions occurred
in the Trojan swarms in very ancient epochs can be expected to survive
until today with very little alterations of their orbital elements,
whereas in the main belt identical groupings would tend to disperse
and be dissipated over relatively short time scales. So, the general
morphology of the stalactite diagrams shown in
Figs.~\ref{fig:stalatL4} and \ref{fig:stalatL5} can be
interpreted in terms of very compact groups which may well be the
survivors of events that occurred in very ancient times. The relatively
small numbers of objects can also be a consequence of the fact that
the smallest objects at this heliocentric distance have not yet been
discovered.

It is also possible that for the Jupiter Trojan population our
criteria for family acceptance, which have been defined to be
applicable primarily to the case of main belt asteroids, could be
overly severe. In other words, we cannot rule out the possibility that
some, relatively deep stalactites that do not satisfy our criteria to
be accepted as families, might still be the remnants of ancient
collisional events.

Unfortunately, one of the major tools we have at disposal to test
similar hypotheses among main belt astreroids, namely the evidence
coming from spectroscopic and/or polarimetric evidence, are much less
powerful a priori when applied to the Jupiter Trojans, which are quite
homogeneous in terms of spectral properties and sufficiently faint to
be a challenge for polarimetric measurements.

\begin{figure}[h!]
\figfigincl{12 cm}{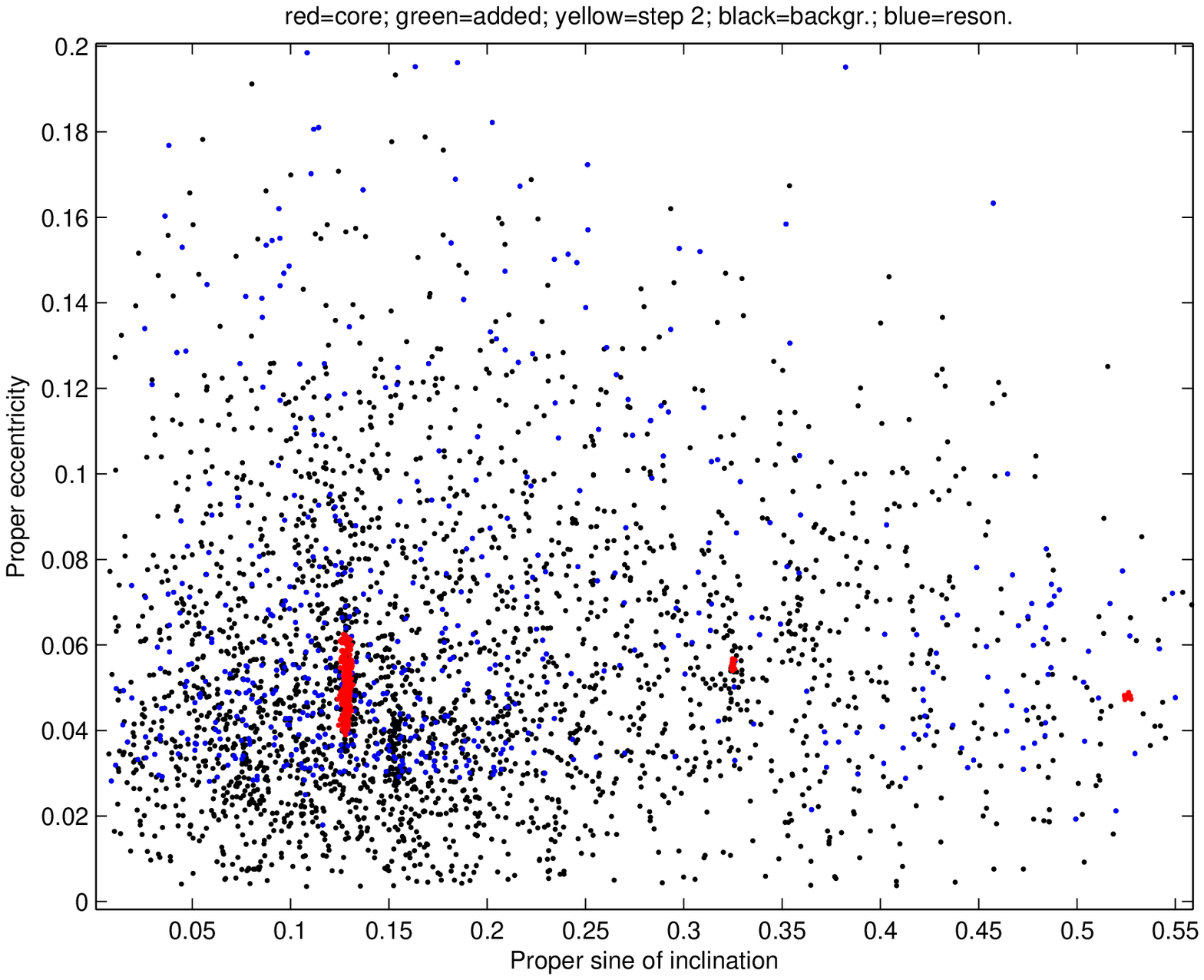}{Proper eccetricity versus sine of
  inclination plot for the L4 Jupiter Trojan swarm.}
\end{figure}

\subsubsection*{The Eurybates family}

Somewhat less compact, but by far the most populous group in the L4 Trojan
swarm is the family of (3548) Eurybates. It includes $172$ members,
$41$ of them being multi-opposition objects.

Strictly linked to Eurybates are 8060 and 222861, two small groupings,
consisting of $25$ members, including $5$ multi-opposition objects,
and $11$ members ($3$ multi-opposition), respectively. These families
are very close to Eurybates, and join it already at the level of $40$
m/s, just $10$ m/s above the level for statistical significance.

The family of (3548) Eurybates is located in the tip of a larger
stalactite branch well visible in Fig.~\ref{fig:stalatL4}.  This
also corresponds to the most evident grouping with proper sine of
inclination about $0.13$ and proper eccentricity around $0.05$
in the plot shown in Fig.~\ref{fig:L4sinie}.  

We have used the same calibration method of Section~\ref{s:hilda} to
get the right order of magnitudes of the secular effects in all proper
elements. On $\Delta a$ the secular effect of Yarkovsky is $\sim
10^{-5}$ of the corresponding effect in $a$ outside the $1/1$
resonance, thus dispersion in $\Delta a$ is negligible even over the
age of the Solar System. For proper $e$ the ratio of $de/dt$ with the
non-resonant $da/dt$ is $3\times 10^{-3}$, for proper $\sin{I}$ it is
$4.6\times 10^{-3}$. Thus, although it is possible to plot the family
members in either a $(e, 1/D)$ plane or in a $(\sin{I}, 1/D)$ plane
and measure slopes of the V-shapes, the values of the ages which would
result from the hypothesis that these V-shapes describe a Yarkovsky
effect are by far too large: the smallest one resulting from the use of
$(\sin{I}, 1/D)$ is $14$ Gy, little more than the age of the universe.

This implies that the distribution of proper elements of the family
members is essentially only due to the original distribution of
relative velocities of the fragments, with very minor contribution
from any non-gravitational perturbation. Thus we cannot estimate the
ages of the Trojan families, but we can assess the distribution of
the original velocities, something which is not possible in the
asteroid main belt (apart from the case of recent families). 

As a check of this approach, we have computed the standard deviation
of the proper elements of the members of 3548, and found $0.0048$ for
proper $e$, $0.0015$ for proper $\sin{I}$ and $0.0107$ au for $\Delta
a$, corresponding to relative velocities after escaping from the
gravitational well of the parent body of $62$, $20$ and $13$ m/s,
respectively.  The parent body can be estimated from the family volume
to have had $D=93$ km, thus the escape velocity can be roughly
estimated (we have very poor density data) at about $100$ m/s.  Thus
the values of the relative velocities obtained from the present values
of the proper elements are fully compatible with an original velocity
field. 

\subsubsection*{High inclination Trojan families}

The families of (624) Hektor and of (9799) 1996 RJ, with $15$ objects
($6$ multiopposition) and $13$ ($6$ multi-opposition), respectively,
correspond to two extremely compact, deep and well isolated stalactite
branches which are evident at the left and rigth ends of the
stalactite diagram shown in Fig.~\ref{fig:stalatL4}.  

Both families have relatively high inclination ($\sin(I) \simeq 0.33$
for 624, $\sin(I) \simeq 0.53$ for 9799). Their extremely compact
structure is clearly visible in the sine of inclination - eccentricity
plot (Fig.~\ref{fig:L4sinie}) where they correspond to very compact
clusters ressembling big blots in the middle-right part of the Figure,
at proper eccentricity values slightly above and slightly below $0.05$,
respectively.

By the same argument used for 3548, the distributions of proper
elements should reflect the original field of relative velocities
after the exit from the sphere of influence of the parent body: both
these families are of cratering type. (624) Hector is a very large,
oddly shaped body (suspect contact binary) with a satellite. The RMS
of proper elements differences, with respect to those of the parent
body (624), are $0.0042$ au for $\Delta a$, $0.0009$ for $e$, $0.0006$
for $\sin{I}$, corresponding to velocities after escape of $5, 11$ and
$8$ m/s respectively. These numbers are so much smaller than the
escape velocity from (624) Hektor, which is about $130$ m/s, to point
out to a marginal escape from a cratering event, possibly the same
resulting in the formation of a satellite and also of the contact binary. 

The same argument can be applied to the family of (9799) 1996 RJ,
which is also of cratering type, with RMS of differences in proper
elements corresponding to $4, 6$, and $15$ m/s for $\Delta a, e$, and
$\sin{I}$, respectively. All the three examples analysed above appear
to confirm that the Trojan asteroid families are \textbf{fossil}
families, frozen at their proper elements which are essentially
determined by the original ejection velocity field. On the other hand,
these velocity fields are remarkably different in cratering
vs. fragmentation cases.

Looking at the same Figure, we also note that few other, less compact
groupings are present, but they do not correspond to acceptable
families according to our HCM criteria. As mentioned above, this might
be a consequence of applying overly conservative criteria of
identification in a region in which families tend to last over very
long timescales without relevant erosion.  Some of these groups might
``grow'' in the future, and become acceptable, when the magnitude
completeness limit for the Trojans will improve. A good example may
become a family of (2148) Epeios, proposed by \cite{vinogradova}, which
can be found by relaxing the HCM significance criteria.

\subsection{2/1 resonance: the Griqua region}

\begin{figure}[h!]
\figfigincl{12cm}{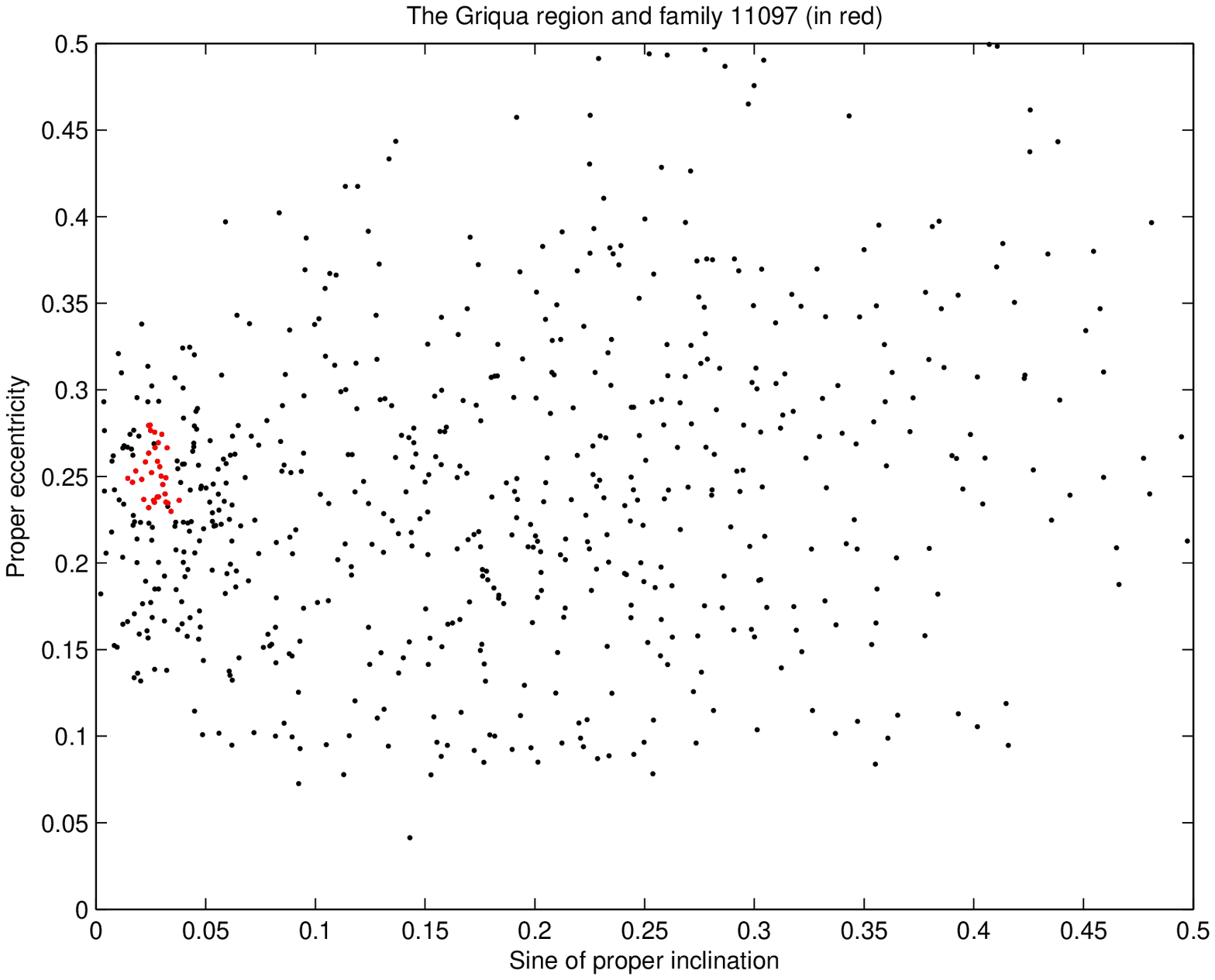}{The Griqua region projected on the
  proper $(\sin{I}, e)$ plane. The red points are the family
  of (11097) 1994 UD1.}
\end{figure}

The Griquas are a comparatively small ($649$ asteroids) population of
asteroids confined in the 2/1 mean motion resonance with Jupiter, in a
tight range of ``semi-proper'' semimajor axis: $3.26<a<3.28$. By the
same argument used for Hildas, the computation of true resonant proper
elements is not necessary. Griquas are fairly spread in proper
eccentricity and sine of inclination, between $0$ and $0.55$ and
between $0$ and $0.65$, respectively, see Figure~\ref{fig:griqua_ie_fam}
\footnote{The name of this population is taken from the lowest-numbered
asteroid present in this region, (1326) Griqua.}.

The classification of Paper I did not find any statistically
significant family in this region. However, the statistics might have
been distorted by merging the sparsely populated $2/1$ region with the
very populated Zone 4, with $a$ above the $5/2$ gap. Thus we have
rerun a dedicated classification for the Griquas only: of course the
quasi-random background became sparse, thus allowing to detect smaller 
families. 

\begin{figure}[t]
\figfigincl{10 cm}{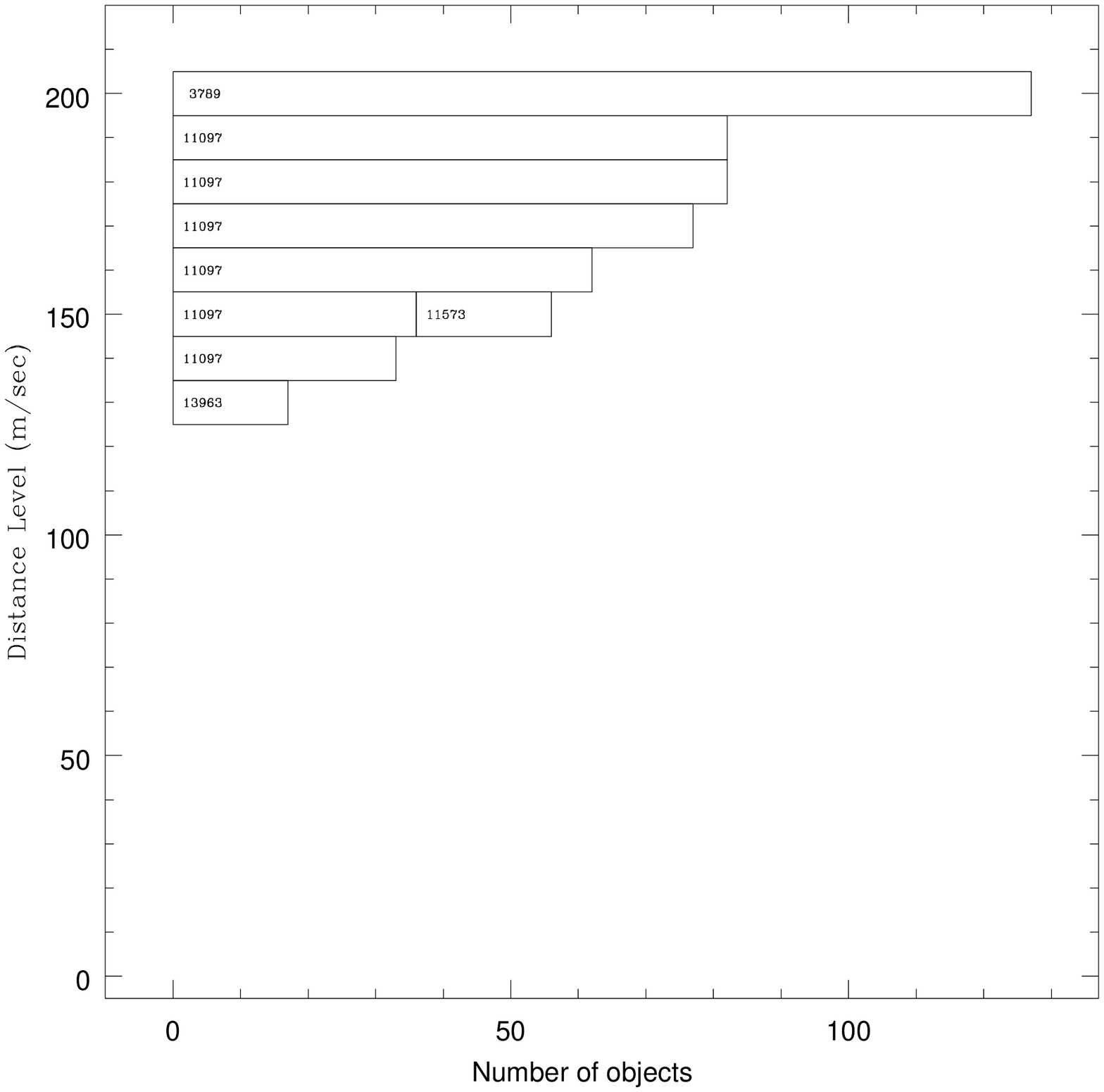}{Stalactite diagram for the Griquas 
population. The minimum number of objects to form an acceptable 
group is $N_{lim} = 16$. The distance level at which families are
identified is $140$ m/s.}
\end{figure}

In this region, we adopt for $N_{lim}$ a value of $16$. The stalactite diagram 
is shown in Fig.~\ref{fig:stalatGriquas_nlim16}. The Quasi Random Level for 
the Griquas is found to be $QRL = 150$ m/s, implying that the distance
level to identify families is $140$ m/s.
Only one such group is present, the 11097 family, which has $33$
members, including $2$ multiopposition objects.  It corresponds to a
group that is apparent in Figure~\ref{fig:griqua_ie_fam}, where it is
located at very low proper inclination.  This is a small family,
which deserves further study, also to be confirmed by observation of a
homogenous composition.


\section{Families affected by secular resonances}
\label{s:secres}

There are in our classification five families with enough membership
to be suitable for age estimation, but affected in significant way by
the secular resonances, in which either all or a good portion of the
members are locked; for a map of the secular resonances in the main
belt, see \cite{propel1994}[Fig. 7-11]. To provide a theory of the
long term dynamical evolution of asteroids affected by both secular
resonances and Yarkovsky effect is beyond the scope of this paper; some
cases have been discussed in the literature \citep{bottke2001,
  voketal06eos}.

Our goal for this paper is just to estimate ages, thus we are going to
use an empirical approach very similar to the one we have used for the
Hildas: by numerical integration we perform a calibration for the
Yarkovsky effect inside the resonance versus the Yarkovsky effect
which would occur to the same asteroid if the resonance was not there.
In this way we solve all five cases, three with either
all or a good portion of the members locked in the resonance
(families 5, 363, 945), and two (families 283, 686) which have only a minor
portion of members affected by a secular resonance, such
that the V-shape in $(a, 1/D)$ can not change by the action of the
resonance.
 
\subsection{The Astraea family}

To understand the shape in proper elements space of the family of
(5)~Astraea is a complex problem, both because almost all its members
are locked in the $g+g5-2\,g_6$ nonlinear secular resonance (with
values $|g+g5-2\,g_6|<0.5$ arcsec/y), and because the proper $e$ has a
large spread, up to $0.236$, a value such that it must be due to a
large number of mean motion resonances affecting the stability of the
proper elements over very long time spans. Note that many of
these mean motion resonances are with inner planets, thus they are not
included in the computation of the proper elements for the asteroids
of the outer main belt ($a>2.5$ au) \cite{knemil2000}. In the case of
the family 5, with lowest value of proper $a=2.55$ au and with
comparatively high eccentricity, to include the inner planets in the
dynamical model, as in \cite{knemil2003}, could improve the results.

Physical observations available for this family are few, because it 
originated by a cratering, thus has especially small members: $77\%$
of the members have $D<2$ km. WISE albedo data \citep{masiero2011}
with nominal $S/N>3$ are available for only $6.5\%$ of the family members; out
of these, $31\%$ of the albedo values are $<0.1$, that is they
indicate interlopers, since (5) has IRAS albedo \citep{tedesco2002}
$0.227\pm 0.027$. This fraction of interlopers is somewhat high, but
not extraordinarily so \citep{miglio, radovic}: it could have been affected by
the ``compression'' resulting from the standard computation of proper
elements which does not specifically account for the secular
resonance. Because this compression applies to both the real family
family and the background, this can explain a larger fraction of
interlopers.  Anyway, this comparatively large fraction of interlopers
is a fact and we need to take it into account in our age estimation
procedure.

\begin{figure}[t]
\figfigincl{12 cm}{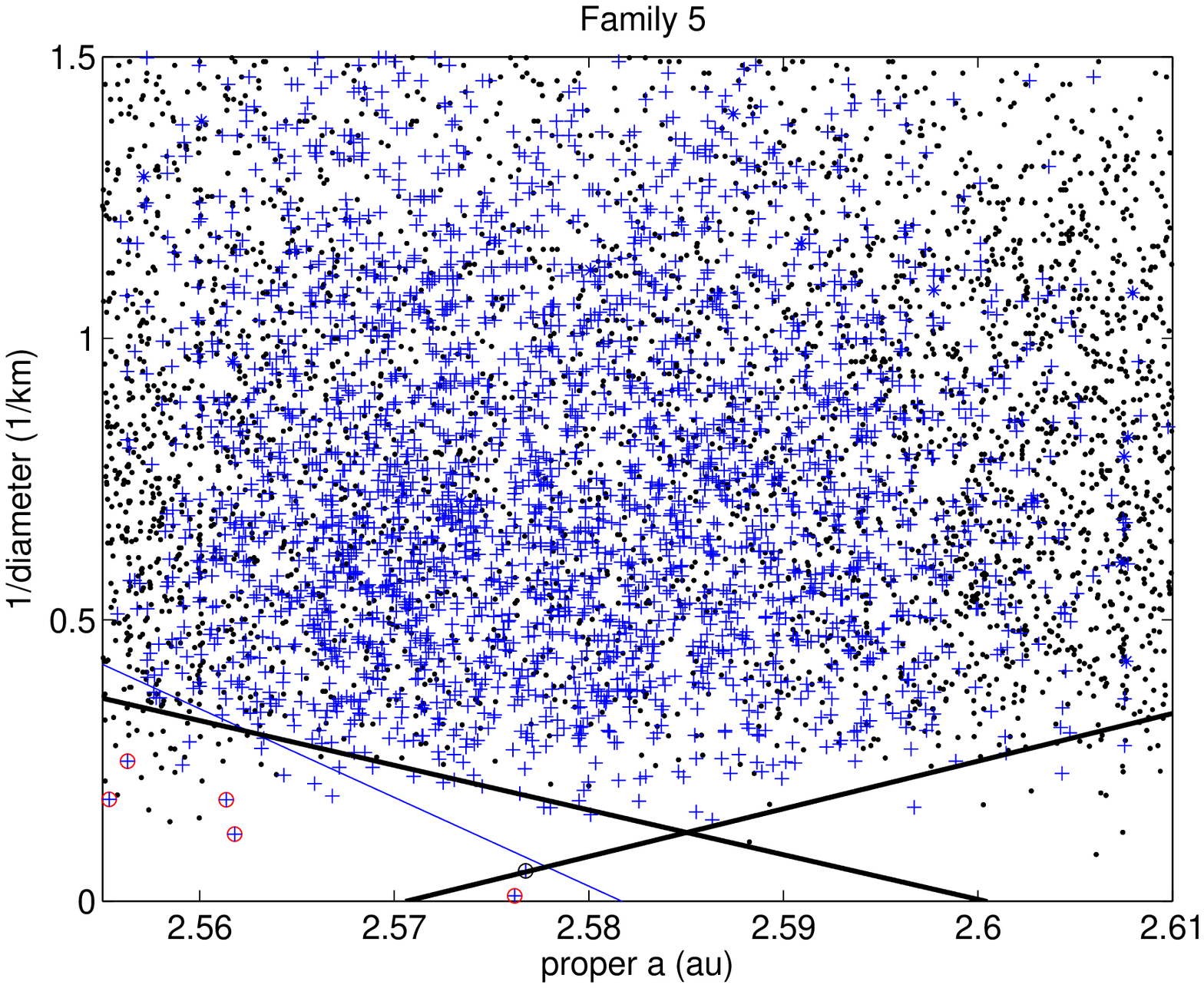}{V-shape of the family of (5) Astraea,
  whose members are locked in the $g+g_5-2\,g_6$ nonlinear secular
  resonance. (5) Astrea is not included in the slope fit because the
  family is of cratering type. Blue crosses are family members, black
  dots are background objects.}
\end{figure}

If we analyse the V-shape in the $(a, 1/D)$ plane
(Figure~\ref{fig:5_vshapea}) we see that the computations of the slope
are possible, although the IN side is affected by a group of four
members too large for their position, near the inner edge of the
family: e.g., (4700) Carusi has an estimated $D>8$~km. Near the center
of the family, (1044) Teutonia has an IRAS estimated $D=15$~km, which
is incompatible with the cratering origin of the family. Moreover, the
fit for the IN side was very poor, to the point that the inverse slope
was barely significant.

We have therefore decided to discard a priori
4 interlopers, marked with a red circle in Figure~\ref{fig:5_vshapea};
then also (1044) Teutonia was automatically discarded as outlier in the
fit. After these removals, the family contains no member with $D>7$~km
and the fit has been much improved, being compatible with a single
collision origin and with a similar uncertainty on the two sides.  Our
choice of four members interpreted as interlopers is plausible also
because of the large fraction of interlopers in general in the family.

\begin{figure}[t]
\figfigincl{12 cm}{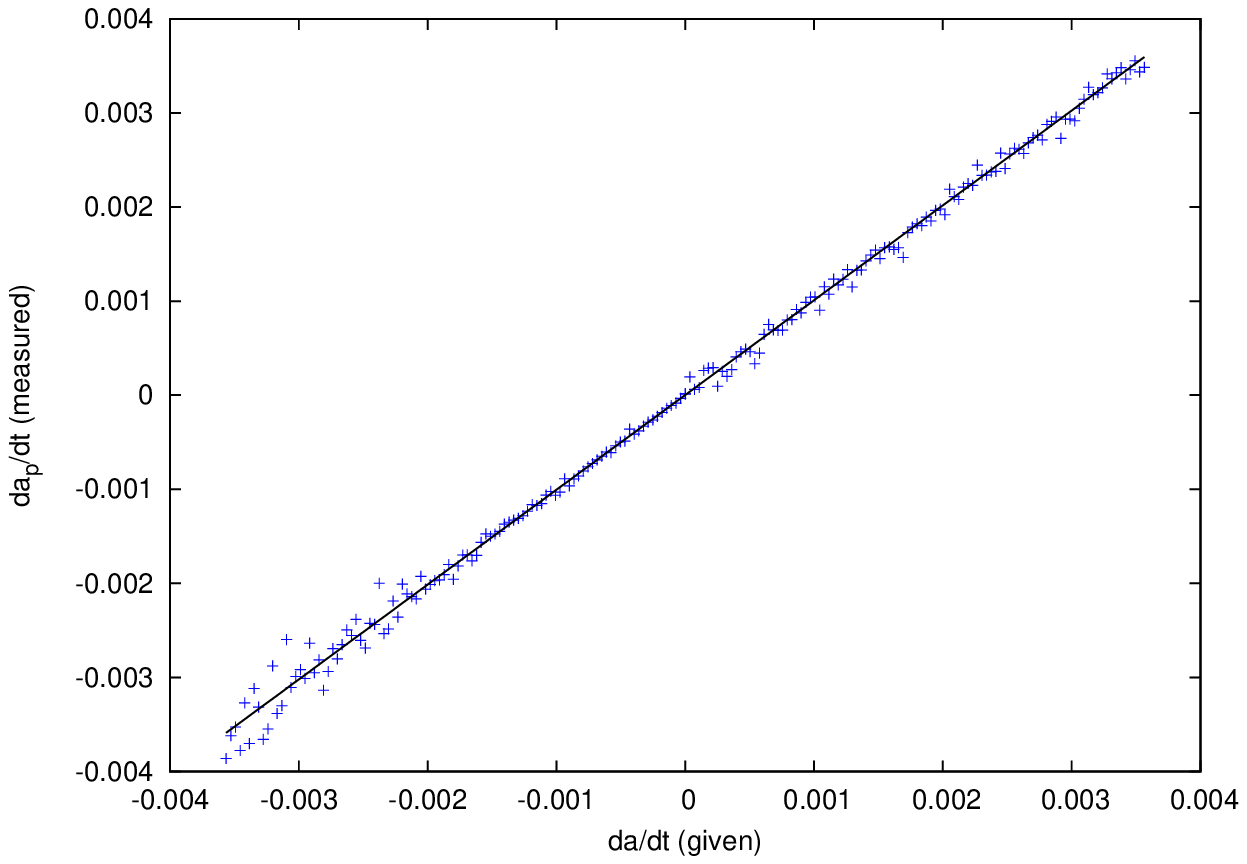}{Yarkovsky calibration for the family
  of (5) Astraea. Slightly larger dispersion of the fit points at the
  far negative side of the regression line is due to the mean motion
  resonances.}
\end{figure}

Before interpreting the inverse slopes of the V-shape as measures of
age, we need to perform a calibration of the Yarkovsky effect taking
specifically into account the dynamical environment of the family.
For this purpose we have performed a numerical integration for $10$
My, with inner planets included, for 200 clones of (5) Astraea. The
right hand side of the equations of motion included Yarkovsky
accelerations corresponding, for a non-resonant objects, to values of
secular $da/dt$ in the range $-3.6\times 10^{-9}<da/dt(NR)<3.6\times
10^{-9}$ au/y. After computing for each clone 9 sets of proper
elements (as in Section~\ref{s:schubart}), we computed for each clone
the ``resonant'' $da/dt(RES)$ drift, then plotted the result in
Figure~\ref{fig:5_calib} with the fit to a line which has slope
$1.008$. In other words:
\[
\frac{da}{dt}(RES)= 1.008 \cdot \frac{da}{dt}(NR)\ ;
\]
meaning that the amount of secular change in proper $a$ due to
Yarkovsky is not significantly affected by the secular resonance.

We can therefore conclude on the age as reported in
Table~\ref{tab:tableages_dadt}: $339 \pm 104$ for the IN side, and
$319 \pm 98$ for OUT, in My; the age of family 5 is old, but far from
ancient, and represents another very good example of cratering,
significantly younger than Vesta.

\subsection{The Padua family}

Our classification includes a dynamical family of (110) Lydia, whith
almost all members locked in the $g+g_5-2\,g_6$ nonlinear secular
resonance (the same as for family 5).  However, (110) has WISE albedo
$=0.17\pm 0.04$ while $90\%$ of the members having WISE data (with
$S/N>3$) have albedo $<0.1$. Thus (110) Lydia is a very likely
interloper and the family namesake should be (363) Padua. We applied
this change in the tables, starting from Table~\ref{tab:tableslopesa}.

\begin{figure}[t]
\figfigincl{12 cm}{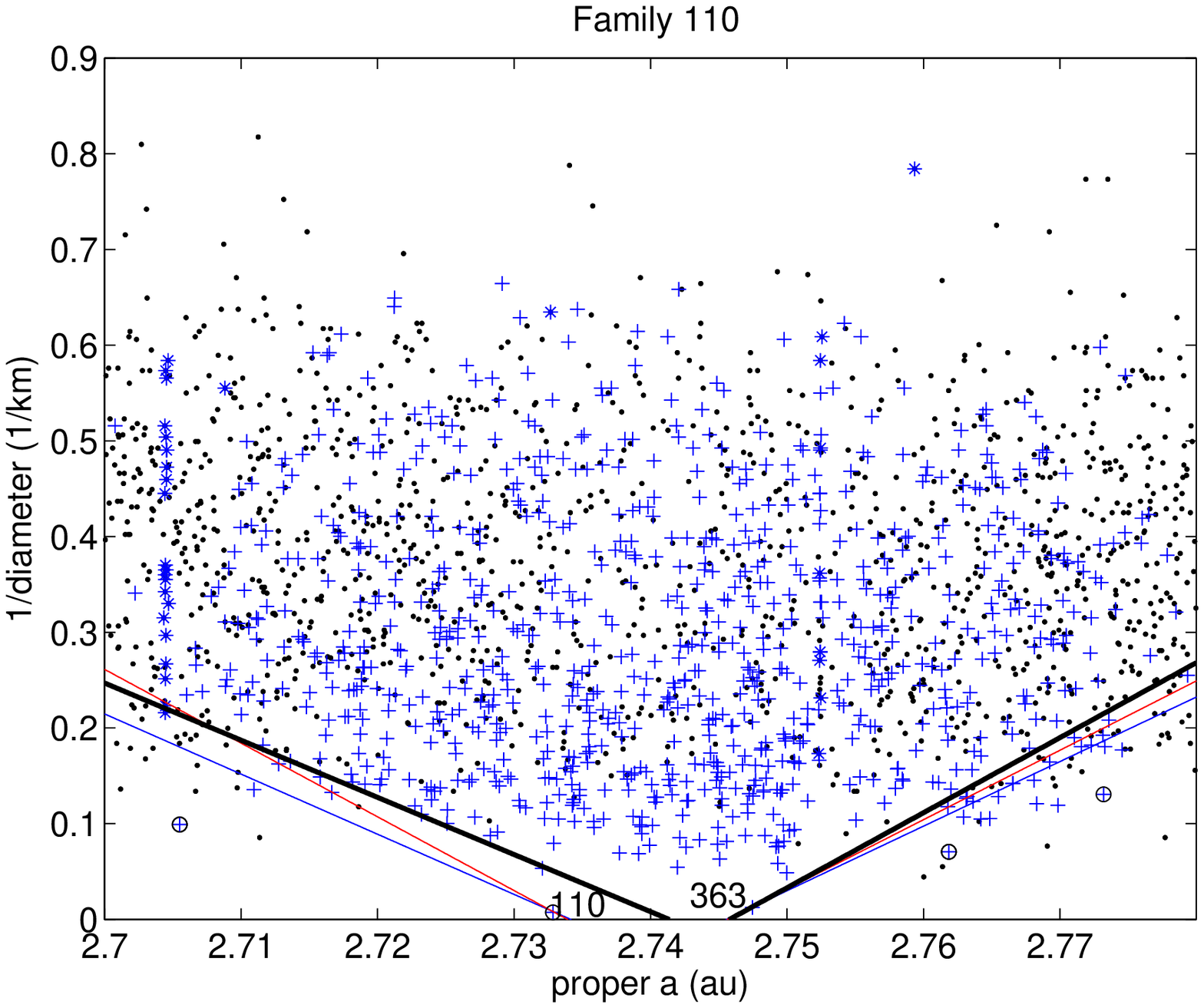}{V-shape of the family of (363) Padua;
  note that (110) Lydia is also an outlier in the slope fit,
  confirming its exclusion from the collisional family by physical
  properties.}
\end{figure}

The V-shape of Figure~\ref{fig:363_vshapea}, besides confirming the
need to remove (110) from the family list, gives good fits for the
inverse slopes with compatible values, suggesting a single
age\footnote{The slopes of the V-shape of family 1726 are $-0.199\pm
  0.028$ on the IN side and $0.192\pm 0.025$ on the OUT side, thus the
  value obtained for family 363 implies that the two families are not
  from the same collision, thus confirming the choice made in Paper IV
  of not merging these two.}.

In the same way as we did for family 5, we have
computed\footnote{Given the larger value of proper $a$, we
  have performed the 10 My numerical integration with a model
  including this time only the outer planets.} the calibration for the secular
$da/dt(RES)$ inside the resonance as a function of the one for
$da/dt(NR)$ outside the resonance, and we got a figure very much like
Figure~\ref{fig:5_calib}, with a fit slope of $0.985$ and some
noise for positive values of $da/dt(NR)$, presumably due to a number
of mean motion resonances, including 3J-1S-1A. We do not think that
the difference between the calibration $1.006$ of the previous case
and this one is significant, anyway it gives a negligible contribution
to the relative uncertainty of the calibration.

After calibration, the ages reported in Table~\ref{tab:tableages_dadt}
indicate a family at the low end of the ``old'' range: $284 \pm 73$
for the IN, and $219 \pm 48$ for the OUT side, in My, compatible with
a single collisional origin of fragmentation type, although with a
largest remnant (363) containing as much as $75\%$ of the family volume.

\subsection{The Barcelona family}

The family of (945) Barcelona has about $2/3$ of the members (the
portion of the family with proper $e>0.23$) strongly affected by the
resonance between the perihelion precession of the asteroid $g$ and
the one of Jupiter $g_5$, with $|g-g_5|<2$ arcsec/y. Another
unidentified nonlinear secular resonance is also relevant for the dynamics
of most members. 


\begin{figure}[h!]
\figfigincl{12 cm}{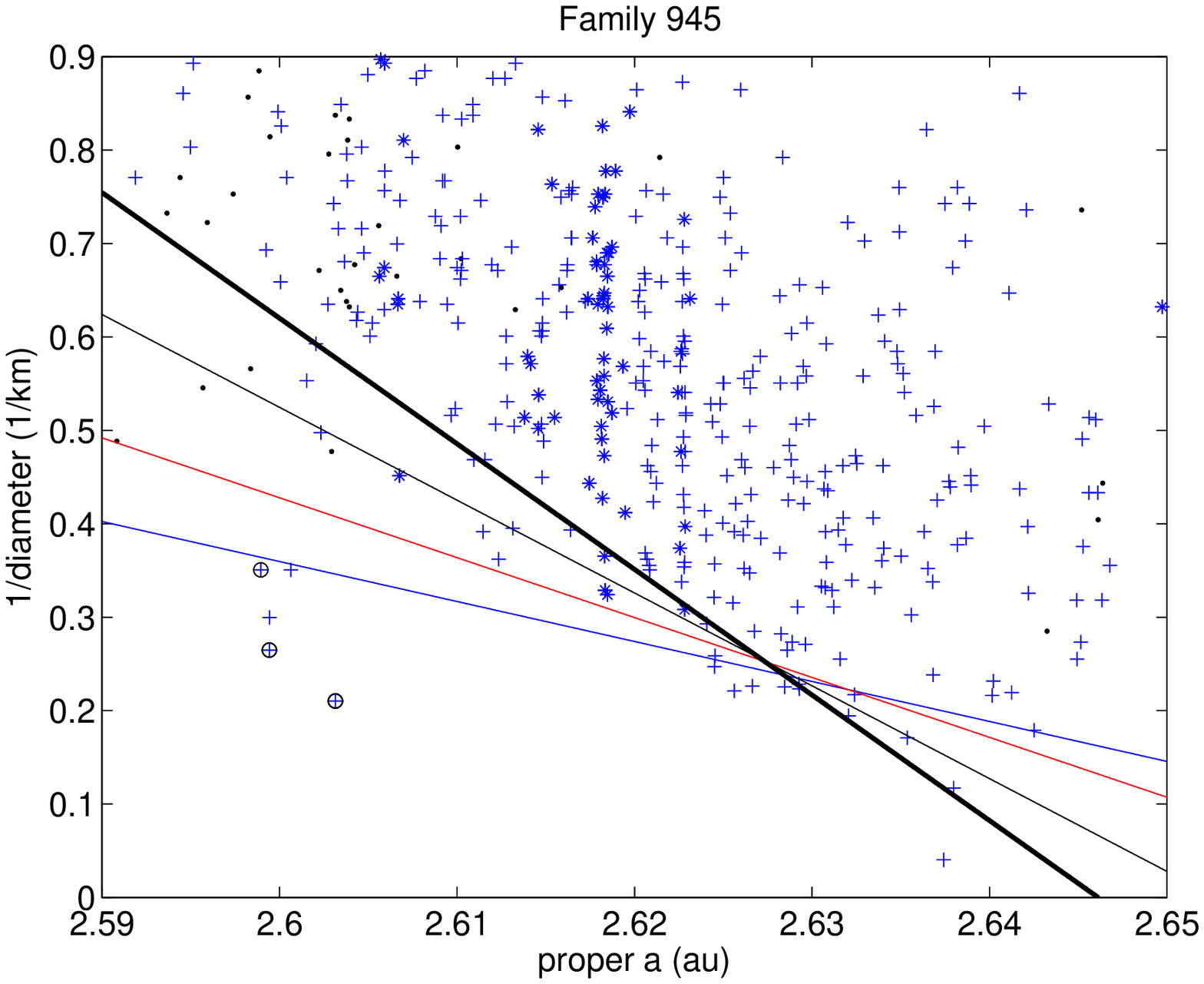}{One-sided V-shape of the family of
  (945) Barcelona.  The gap on the right, where the family terminates
  on the OUT side, corresponds to the $11/4$ J resonance at $a\simeq
  2.649$.}
\end{figure}

The family is bound on the OUT side by the $11/4$ mean resonance with
Jupiter, thus its V-shape (Figure~\ref{fig:945_vshapea}) has only the
IN side.  The IN side fit is good, with some outliers, allowing to
estimate the inverse slope with good accuracy.  

We confirm our approach of a purely numerical calibration of
$da/dt(RES)$, not only because a theory of the evolution under
Yarkovsky effect plus two overlapping secular resonances is out of the
scope of the present paper, but as it also appears exceedingly
difficult.  The same calibration as in the case of family 363 gives a
slope of $da/dt(RES)$ versus $da/dt(NR)$ of $1.046$. The plot shows
much more noise than in the previous two cases, some of which is
associated with the 11/4 resonance: this noise does not affect the
family which has been depleted by the resonance. Still our conclusion
is the same: the effect of the secular resonance on the rate of
Yarkovsky is small, to the point of not being significant with respect
to the relative calibration uncertainty.

The age of $203\pm 56$~My is again at the low end of the ``old'' range, the
family is of fragmentation type but with a comparatively large remnant
(945) with $73\%$ of the family volume.


\subsection{The Gersuind family}

Our classification contains a dynamical family 194, with $408$
members.  However, the family namesake (194) Prokne has albedo
$0.052\pm 0.015$ (WISE) thus it does not belong to the family, which,
after removal of members with albedo $<0.08$ and $>0.25$, has a mean
albedo $0.145\pm 0.037$ . The low albedo of (194) has been confirmed
with recent polarimetric measurements by \cite{CAPS}.

After the removal of such a large interloper, (686) Gersuind remains
the lowest numbered member, with albedo $0.142\pm 0.037$, thus the
family has to be called family 686; we applied this change in the
tables, starting from Table~\ref{tab:tableslopesa}. A family of
(686)~Gersuind, without (194), had already been proposed by
\cite{bojan_highi}.

As afore mentioned, this family is only partially affected by the
$s-s6-g5+g6$ nonlinear secular resonance, with only some $14\%$ of
members in the OUT side having the corresponding frequency
$|s-s6-g5+g6|<0.5$ arcsec/y.

\begin{figure}[h!]
\figfigincl{12 cm}{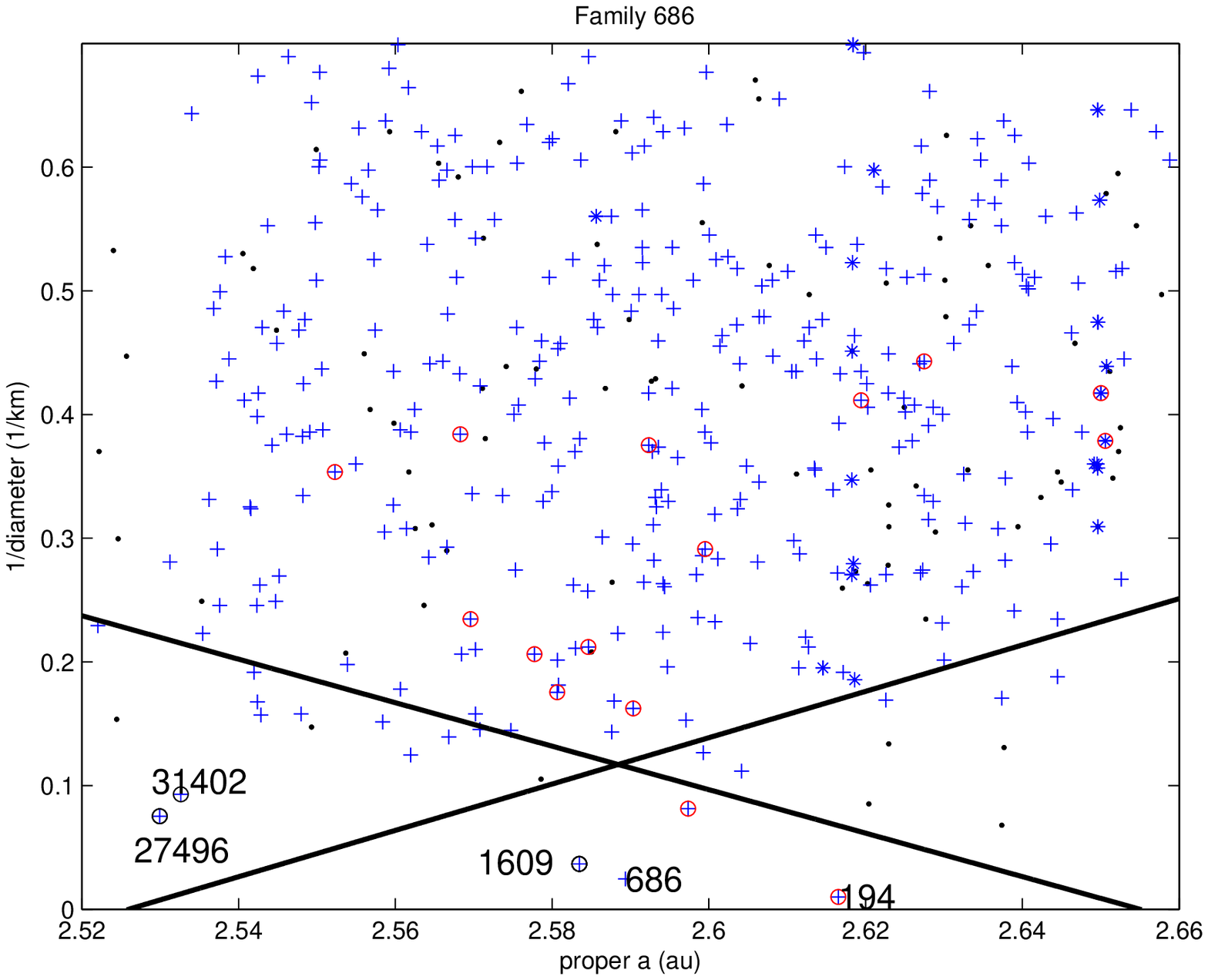}{V-shape of the family of (686)
  Gersuind. The asteroids marked with a red circle are interlopers
  discarded by albedo, the ones with a black circle are removed from
  the fit as outliers.}
\end{figure}

The V-shape of the family 686 (Figure~\ref{fig:686_vshapea}), with
(194) and other members with incompatible albedo removed, still
results in a poor fit for the IN side, with $1/S=-0.569\pm 0.322$, but
in a better one for the OUT side $0.534\pm 0.138$. After calibration,
which yielded regression line slope of 1.003, the estimated ages are
$1490\pm 843$ and $1436\pm 469$~My, with the OUT value much more
significant.

Note that the members (1609), (27496) and (31402) are removed as
outliers from the slope fit: since there are no albedo data their
status as interlopers remains to be confirmed.

\subsection{The Emma family}

The family of (283) Emma contains dark asteroids, with few known
higher albedo interlopers: the mean WISE albedo is $\sim 0.05$,
(283) itself has an IRAS albedo $0.026$. A small fraction (about $8\%$) of
its members are affected by the nonlinear secular resonance
$g+s-g6-s6$, which opens a kind of gap between two parts of the
family, well visible in the projection on the proper $(a,\sin{I})$
plane, for low values of $a$.

The V-shape (Figure~\ref{fig:283_vshapea}) results in a good fit on
the IN side with $1/S=-0.165\pm 0.019$, thus it does not appear to be
much affected by the secular resonance. To the contrary, on the OUT
side the fit is poor: $0.355\pm 0.112$. After calibration performed on
the second largest family member (32931) Ferioli (regression line
slope 1.010), the age estimates are $290\pm 67$ on IN side, and
$628\pm 234$ My on the OUT side. The family is of cratering type:
assuming same albedo for all, the volume of the family members without
(283) is about $9\%$ of the total. Thus the occurence of two separate
cratering events is a reasonable hypothesis, which is consistent with
a double jet shape as shown in the proper $(a,\sin{I})$ projection.

\begin{figure}[t]
\figfigincl{12 cm}{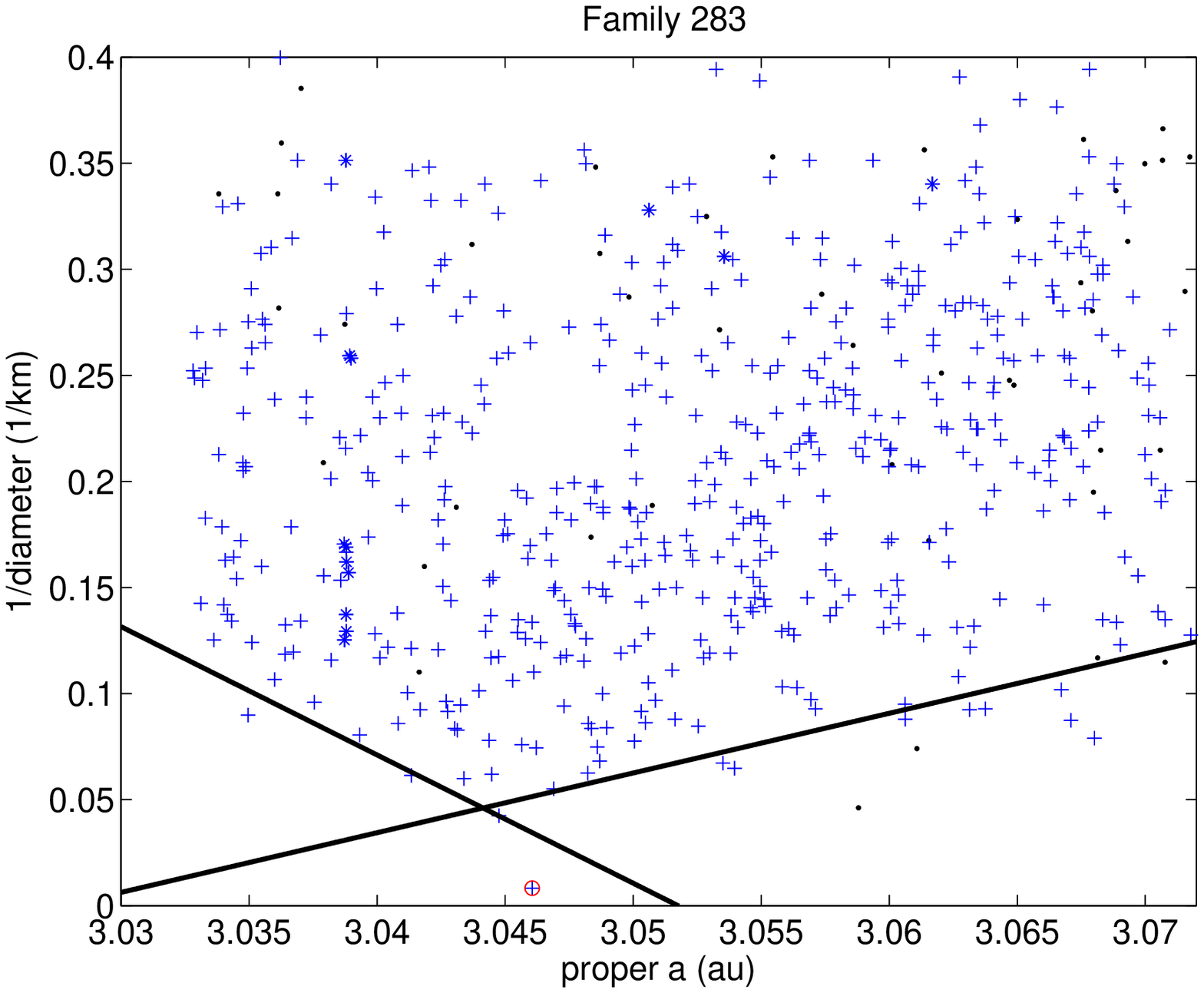}{V-shape of the family of (283) Emma;
  the two slopes are ostensibly different. Note that the family is of
  cratering type, thus the parent body (283) Emma is excluded from the
  fit (marked with red circle), as well as from calibration.}
\end{figure}

\subsection{The Phocaea family}

\begin{figure}[t]
\figfigincl{12 cm}{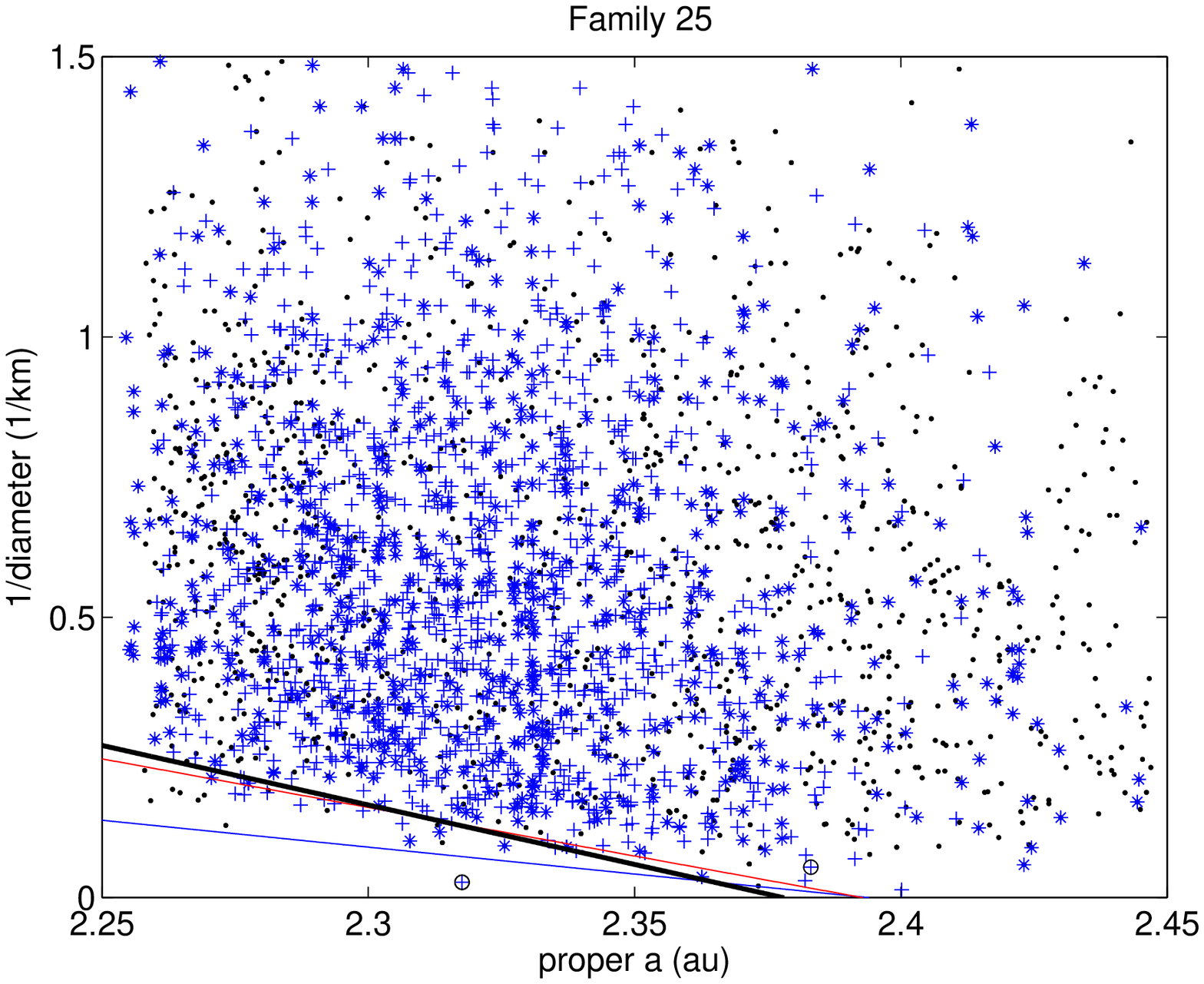}{V-shape of the family of (25)
  Phocaea. The circled cross is (326) which is an outlier for the
  V-shape fit as well as an inteloper because of its low albedo.}
\end{figure}

The family of (25) Phocaea has a one-sided V-shape in the $(a, 1/D)$
plane (Figure~\ref{fig:25_vshapea}), with (25) at the high $a$ end.  The
missing part of the V-shape is due to the resonances bordering the
stable region, including the $3/1$ J. The region contains about
$4,000$ asteroids, while the dynamical family 25 contains only $1,405$
of these, thus it represents a substructure whith distinctive number
density inside the stability region. The fit gives $1/S=-0.471\pm
0.047$.

The only problem with the shape of the family as shown in
Figure~\ref{fig:25_vshapea} is that there is obviously a lack of family
members near the value of proper $a$ of (25) Phocea, which occurs
before the high $a$ end of the stability region. This might be
interpreted as a YORP induced central gap \citep{paolicchi}.

$1/3$ of the family members are affected by a nonlinear secular
resonance: they have $|g-s-g_6+s_6|<0.5$ arcsec/y. This fraction is
relevant for the age computation because it may affect the side of the
V-shape, thus we need calibration, which has been done in the same way
by using clones of (323) Brucia (which is locked in the resonance),
but only with negative Yarkovsky drift in proper $a$. The slope of the
$da/dt(RES)$ with respect to the one outside the resonance was found
to be $0.977$, thus once more the calibration does not need to take
into account the resonance. Thus the calibrated age is $1187\pm 319$
My, an ancient family according to the terminology introduced in Paper III.


\section{Irregular families}
\label{s:complex}

By irregular families we mean dynamical families with internal
structures that cannot be interpreted as a simple, single collisional
family; this either prevents the computation of an age, or results in
an age which refers to only a portion of the dynamical family. In some
cases, an interpretation as two or more collisional families
(partially overlapping) is possible, e.g., Vesta, Eunomia,
Agnia. Alternatively, this apparent complexity could be due to a
number of comparatively large interlopers. In other cases the shape of
the family can have different, even more peculiar interpretations. In
this section we shall propose a solution, including an age estimation,
for one of the previously unsolved cases. As explained above, family
of (283) Emma obviously fits to this class, but being also crossed by
the secular resonance, we preferred to deal with it in
Section~\ref{s:secres}.

\subsection{The Adeona family}

\begin{figure}[h!]
\figfigincl{12 cm}{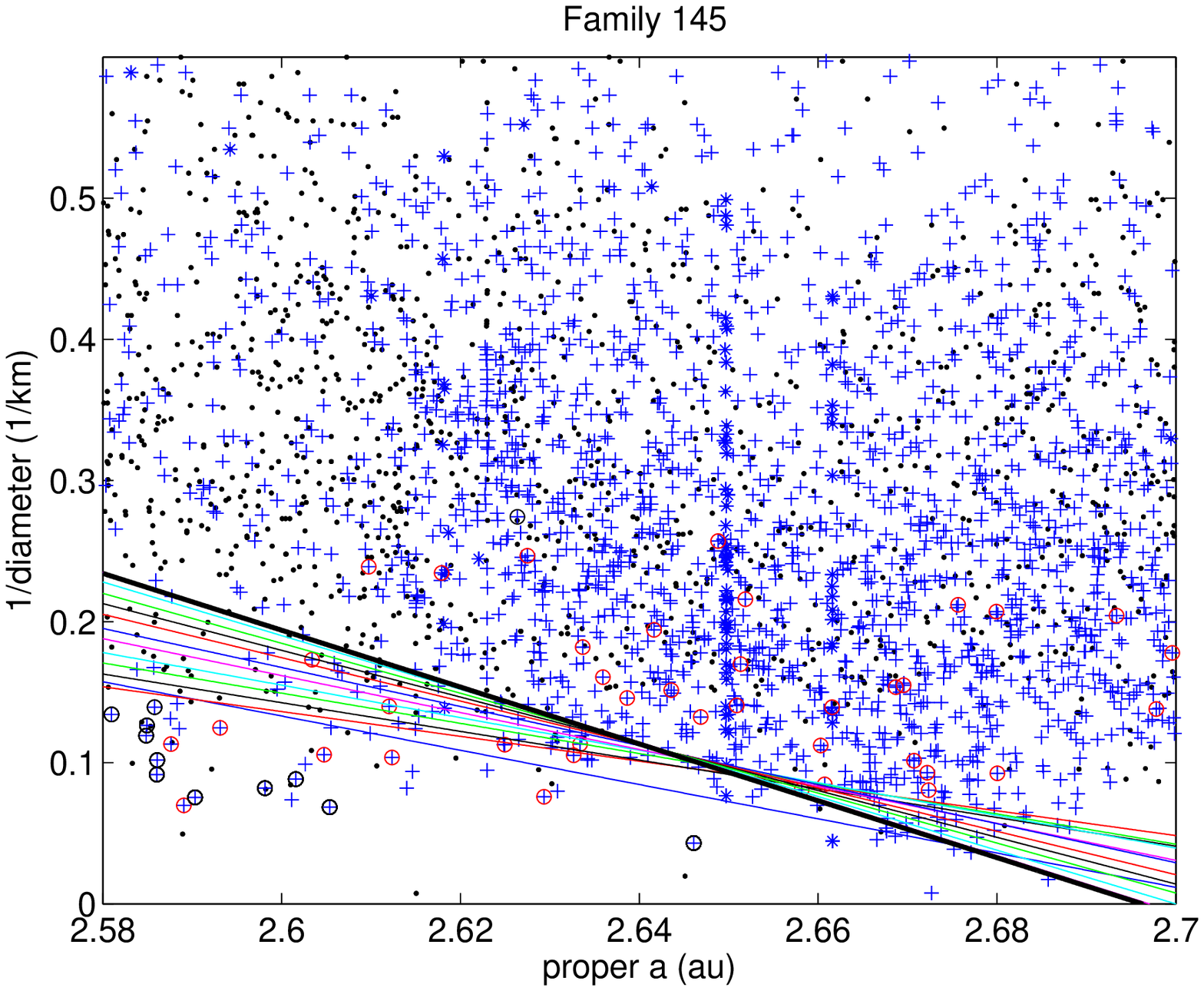}{One-sided V-shape of the family of
  (145) Adeona.  The crosses with red circles correspond to
  interlopers based on physical observations of either albedo or color
  indices; the ones with black circles are outliers removed from the
  V-shape fit. The colored lines represent intermediate steps in the
  fit, due to a large number of outliers being removed in several
  iterations.}
\end{figure}

The family of (145) Adeona has a peculiar shape in the proper $a$,
$1/D$ plane which appears different from the standard V-shape. There
appears to be a common V-shape for a narrow range of values, roughly
$2.63<a<2.70$, while the portion of the family for $2.57<a<2.62$ is
much less dense and does not appear to be delimited by the main
V-shape.

\begin{figure}[t]
\figfigincl{12 cm}{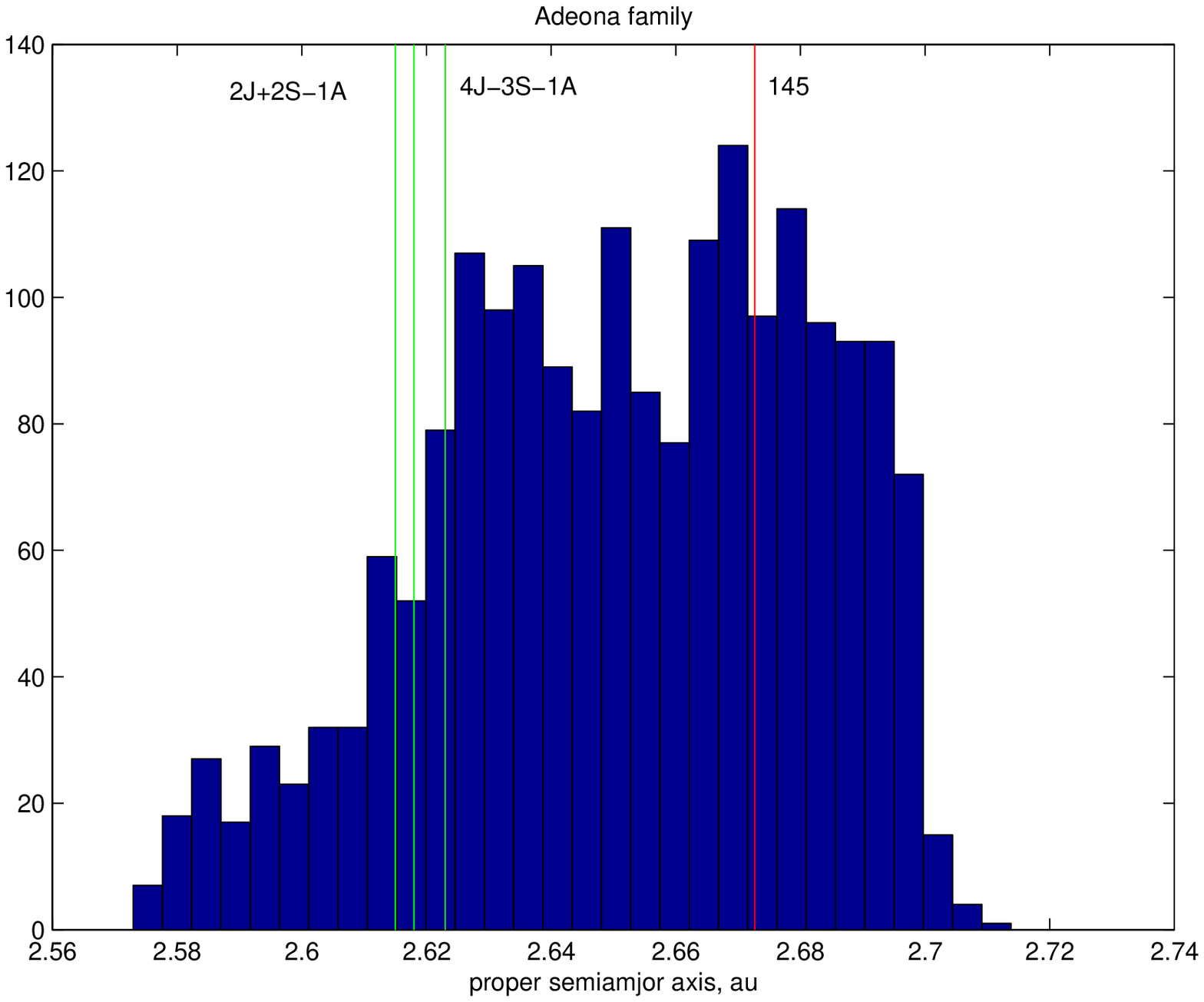}{Histogram of proper $a$ for the family
  of (145) Adeona. The region of higher number density ranges from
  about $2.62$ to $2.70$, and contains (145) Adeona itself, marked by
  the red line. The region of lower number density corresponds to
  values of $a$ below the three resonances marked by the green lines.}
\end{figure}

The histogram of number density as a function of proper $a$ (see
Figure~\ref{fig:145_hista}) shows a sharp decline of dynamical family
membership in the region with $a$ below the three 3-body resonances,
4J-3S-1A at $a\sim 2.623$, 2J+2S-1A at $a\sim 2.615$ (\cite{smirnov}),
and another not identified one at $a\sim 2.618$ au.

This complex structure can have different interpretations: the portion
of the dynamical family for $a<2.615$ could be another, smaller family
with lower number density. The same portion could also be a
continuation of the family of (145) Adeona with decreased number
density, since a large fraction of members could have been removed by
the resonances: then the largest members in the low $a$ region should
be interlopers.

There is, however, a certain indication from physical observations:
the dynamical family contains mostly dark asteroids - $94\%$ with WISE
albedo $<0.1$ - with a few interlopers. The unexplained group of
comparatively large members for $a<2.62$ au contains a fair fraction
of interlopers, characterized either by albedo $>0.1$ or by the color
index $a^*>0$ (see the circled crosses in
Figure~\ref{fig:145_vshapea_in}).

If we just fit the data to a V-shape with the IN side only, we get an
inverse slope $1/S=-0.497\pm 0.058$. As shown in
Figure~\ref{fig:145_vshapea_in}, the number of outliers removed to
obtain this fit is substantial, on top of the interlopers removed by
physical observations. The lack of the OUT side can be due to the
strong $8/3$J resonance at $a\sim 2.704$~au where the family abruptly
ends up.  Thus the age can be estimated as $794\pm 194$ My. As
discussed in Section~\ref{s:obstest}, this result needs to be
confirmed by showing that to discard that many data points is
justified.

\section{Observational tests of families}
\label{s:obstest}

As described in Paper I, our approach to asteroid family analysis is based on
using the proper elements data first to identify the families, then
adding the absolute magnitude data (with complementary information
from the statistics of the albedo data in each family) to estimate
ages, then use other available physical data to refine the results,
e.g., identifying complex familes with more than one collision and
interlopers which can deteriorate some age estimates. However, in many
cases we do not have enough physical data to solve the problem. Thus
we discuss in the following several open problems for which dedicated
observations of asteroid physical properties could decisively
contribute to the solution of some interesting problems.

\subsection{The Hertha complex}

The single dynamical family of (135) Hertha has been known for a long
time to include multiple collisional families, overlapping
in the proper elements space \citep{cellino2001,CellinoAstIII}.
However, we have been able to compute only one age (Paper IV), for the
subfamily of (650) Amalasuntha, which is partly separate in the
$(a,e)$ plane and also distinguished by low albedo and other physical
properties typical of the C taxonomic complex. 
A possible splitting of the low-albedo component of the 
Hertha complex in more than one group, in particular the presence of
an independent Eulalia family \citep{walsh13, dykhuis15} has not been
supported by recent spectroscopic observations by \cite{deleon16}.

The complementary ``bright'' subfamily does not have a recognizable
V-shape. The recently proposed Hertha 1 and 2 families
\citep{dykhuis15} were found by using color index separation based on
SDSS parameters. The presence of three density contrast features is
confirmed by our last classification, in which family 135 has a total
of $15,442$ members\footnote{See the proper $(a,e)$ projection of
  family 135 on our family visualizer at
  http://hamilton.dm.unipi.it/astdys2/Plot/index.html; note the
  bimodality of the bright component shown by the number density
  histogram on the right border of the plot.}; however, the location
of the two ``bright'' subfamilies does not correspond precisely to
what was proposed by \cite{dykhuis15}. Moreover, the role of (135)
Hertha as (core of) parent body has not been established.

Targeted physical observations of several members of family 135, by
selecting asteroids in such a way that they belong to the
separate density features, could provide a robust collisional
interpretation. 

\subsection{The Baptistina complex}

\begin{figure}
\figfigincl{11 cm}{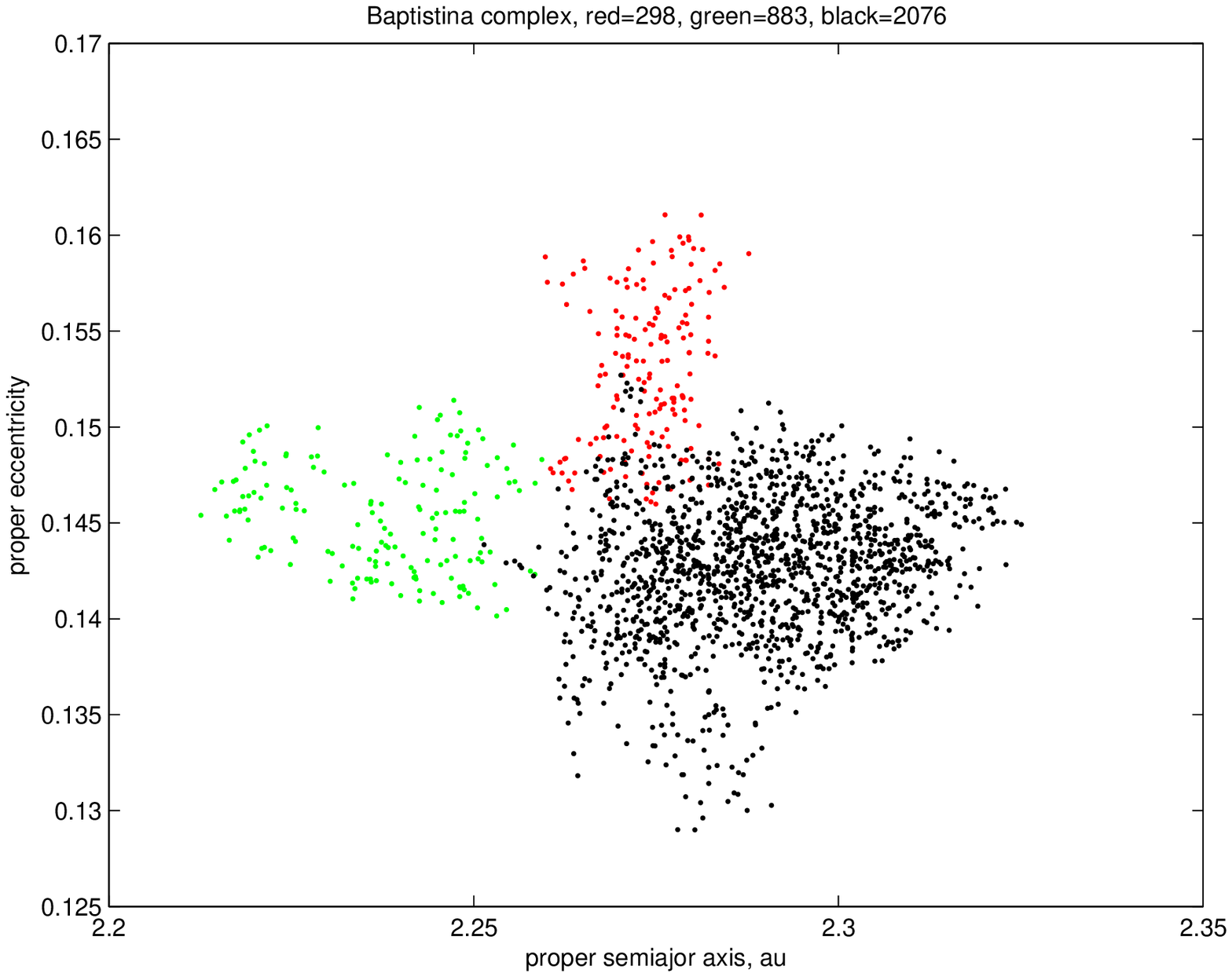}{Distribution in the plane of proper
  $(a,e)$ of the members of families 298 (red), 883 (green), 2076
  (black). There are currently 2 intersections between 298 and 2076
  and 1 intersection between 883 and 2076. The projections on 
 other proper element planes $(a, \sin{I})$ and $(e,\sin{I})$  show a similar complexity.}
\figfigincl{11 cm}{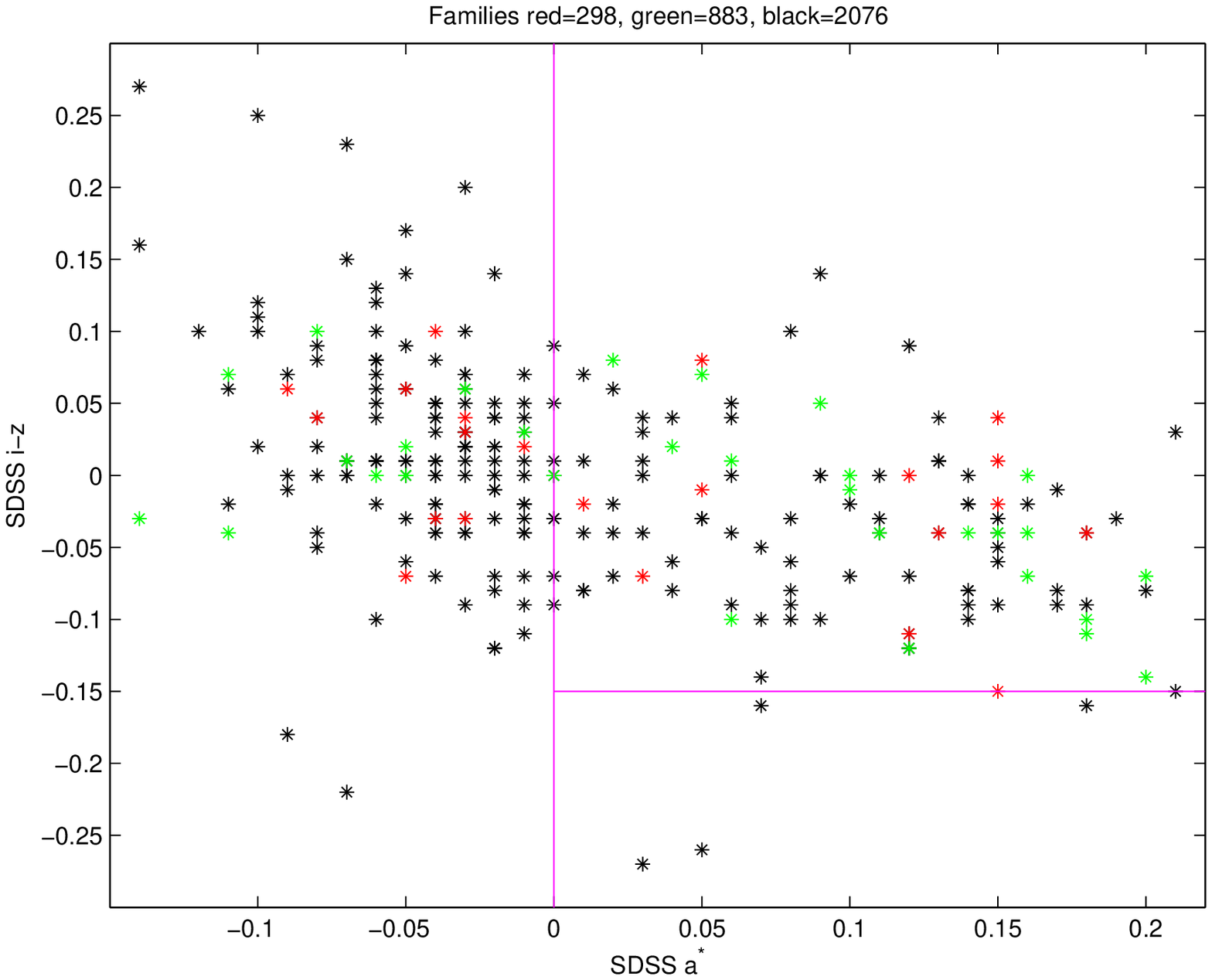}{Distribution in the plane of SDSS
  color indexes $a^*$ and $i-z$ for the members of families 298 (red),
  883 (green), 2076 (black). The magenta lines separate the regions
  mosly occupied by the three taxonomic complexes C (left), S (right
  above), V (right below).}
\end{figure}

For the complex of three dynamical families of (298) Baptistina, (883)
Matterania, and (2076) Levin (see Figure~\ref{fig:baptistina_ae}) we
do not have a collisional interpretation. The three families have few
intersections \footnote{By intersection we mean members in common, see
  Papers I and II.} in the current classification, but we do not see
any way to model them as the result of a single collision. Even an
interpretation in terms of two collisions does not solve all the
problems. The physical data (see Figure~\ref{fig:baptistina_sdss} for
the SDSS data) only make the problem more difficult, because all three
families appear heterogeneous in composition; the same result is
obtained by using WISE albedo data. Therefore we have abstained from
merging these three families, in spite of the intersections, and we
have computed only one age from the one-sided V-shape of 2076 (Paper
III). In the literature there are several ``Baptistina families'' with
rather different membership, conflicting age estimates and conjectures
on composition, all based on the assumption that there is just one
collisional family.

To solve this problem more physical observations are needed; they
should be of high quality and for a large enough set of asteroids in
each of the three families.

\subsection{The Eos family}
\label{s:eos}

The family of (221) Eos is not a problem, in that most of its
membership is well known and a single age has been computed (Paper
IV). However, interlopers with low albedo could belong to another
family formed by cratering of (423) Diotima, which is much larger than
(221). Targeted physical observations in the region near (423) in
proper elements space are required to confirm the existence of this
separate but completely overlapping family.

Another problem is about the family of (179) Klytaemnestra, which is
very close in proper elements space to the family 221 but has no
intersections with it. This family V-shape in $(a,1/D)$ has a consistent
collisional interpretation neither with 1 nor with 2 collisions.  The
only explanation we may propose is that family 179 could be
contaminated by a large number of interlopers (in the low $a$ portion)
transported from family 221, possibly through the secular resonance
$g+s-g_6-s_6$, well known to affect Eos family \citep{milkne92,voketal06eos}.
This could be confirmed by physical observations of two sets
of asteroids, belonging to 221 and 179, respectively, but close in
proper elements space.

\subsection{The Adeona family}

To be able to compute an age for the family of (145) Adeona, we had to
discard a comparatively large number of interlopers, based on the
existing physical data. Additionally, there are many outliers for the
slope fit for which we have no physical data. A precisely targeted
campaign of physical observations of members of the dynamical family
145 with $a< 2.64$ and $5<D<15$ km could either confirm our results,
including the age, or suggest another interpretation, like the
existence of a subfamily.

\subsection{The Gersuind family}

The family of (686) Gersuind, emerged from the dynamical family 194,
may be interesting in connection with the problem of the Barbarians,
asteroids with peculiar polarimetric properties
\citep{cellino06}). Several Barbarians were found in the neighboring
family of (729) Watsonia (\citep{cellino14}). After the removal of the
dark and too bright interlopers from 194, the family 686 has albedo
properties very similar to 729. Thus the members of 686 should be
subjected to investigation with polarimetry.

\subsection{The super-Hilda family}
\label{s:superhilda}

The extended Hilda family we discussed in Section~\ref{s:hilda} is
only a proposal, for which we do not have a statistically significant
confirmation. On the other hand, if it was really there, it would be
especially interesting as one of the oldest families. Given the
difficulty of getting confirmation from physical observations
(essentially all the asteroids in the Hilda region for which data are
available are P types, with low albedo), the only way to test the
``super-Hilda hypothesis'' observationally is to discover more Hildas
in the medium inclination region. It should be possible to
exclude that the very shallow size distribution is due to
observational selection by pushing the complenesss of the catalogs
down to $D= 4$ km. Then this would confirm that the alleged
super-family is indeed strongly eroded, thus very ancient.

\section{Conclusions}
\label{s:conclusions}

In this paper we have attempted to give a collisional model to a
number of families for which the same attempt had previously
failed. Most of these families were either locked in resonances or
anyway significantly affected by resonances, both mean motion and
secular. To estimate an age for the family required in each of these
resonant cases to apply a specific calibration for the Yarkovsky
effect, which in principle could be different in each case. 

For the largest families found in the Hilda region, consisting of
asteroids locked in the 3/2 resonance with Jupiter, the Yarkovsky
effect results in a secular change in eccentricity, thus the V-shape
technique had to be applied in the $(e, 1/D)$ plane. There we found
family 1911 with a good age determination and family 153 of the
\textit{eroded} type, that is which can be seen visually in the
plots in the proper $(\sin{I}, e)$ plane but cannot be confirmed by
the statistical tests of the HCM method. If such a family exists, then
its age must be $>3.5$ Gy, extremely ancient or even primordial, which
is consistent with the hypothesis that this family is depleted to the
point of not having a significant density contrast with the
background.

For the Trojans, that is the asteroids locked in 1/1 resonance with
Jupiter, we are presenting in this paper a new classification which
identifies a number of families by using synthetic proper elements and
a full HCM method. HCM is well established and has been succesfully
applied to the main asteroid belt, but the results in the Trojan
swarms indicate that families there have a very different structure.
Numerical calibrations have shown that the Yarkovsky perturbations are
ineffective in determining secular changes in all proper
elements. This implies that all Trojan families are \textit{fossil}
families, frozen with the original field of relative velocities, which
are small for cratering families, for the fragmentation case somewhat
larger, but still limited to the order of the escape velocity from the
parent body. Thus we find no way to estimate the ages of the families,
while they can be a reliable source of information on the original
velocity field immediately after the collision.

We have found a new family among the Griquas, locked in the 2/1
resonance with Jupiter.

We have analysed 6 large families affected by secular resonances,
mostly the nonlinear ones. We have used a numerical calibration
method, which has shown in all cases that the secular resonances do
not affect significanly the secular change of proper $a$, thus the
V-shape method in the $(a,1/D)$ plane can be used to compute the age
in the standard way (as in Paper III). 

The solution of some of the cases has been possible only by removal of
a number of assumed interlopers. This applies especially to family 145
which otherwise would appear to have a double V-shape. 

In conclusion, with the 10 ages computed in this paper, the situation
is the following: of the 25 families with $>1000$ members, we have
computed at least one age for all but 490, which has a too recent age,
already known, unsuitable for our method. Of the 19 families with
$300<N<1000$ members, excluding 778 which has a too recent age,
already known, there is only one case left without age, namely 179
(see Section~\ref{s:eos}).

Of the 24 families with $100<N<300$ we have computed 6 ages, the
others we believe could only give low reliability results. In this
range only for families with a small range in proper $a$ we can
compute a reliable slope. Thus we can compute only young to medium old
ages ($<200$ My).  As an example, the family of (1222) Tina, with only
$137$ members, has a short $a$ range and appears to have a well
defined V-shape in $(a, 1/D)$, but the inverse slopes on the two sides
are incompatible ($1/S=-0.054\pm 0.007$ IN, $0.032\pm 0.003$ OUT),
thus we should conclude it is the result of two separate
collisions. With so few data points, this does not appear a mature
result, but something to be reanalysed when the number of members is
at least doubled.

Among the problems with family ages we have left open, there are three
complex families which we believe have more ages than we have
estimated: 221, 135, and 298. Two apparently complex cases we believe
have been solved: 145, 283, although some confirmation would be useful
for 145.

Overall, we believe we have completed a useful work, which is based on
the dataset of proper elements we have produced, and by using physical
data only as check (apart from absolute magnitudes used for ages).  We
think it is anyway a good start towards the goal of an absolute
chronology of the main collisions in the asteroid main belt, see in
Figure~\ref{fig:famage_plotnf} all the family ages we have been able
to compute so far (in Papers III, IV, and the present one).  We have
used for the error bars, for the cases where two values IN and OUT are
available and compatible, standard deviations computed by the formulae
from \cite{orbdet}[Section 7.2], applicable to all the cases in which
two least square fits can be merged under the assumption that the
parameters to be determined are the same. This needs to be applied to
the term in the error budget due to the fit of the two slopes, under
the assumption that they are two solutions for the same physical
quantity. Thus  the two fits for the slopes reinforce each other (if
compatible), while the calibrations are affected by the same errors
(e.g., in the density). This more complex formula is an improvement with respect
to what we did in Paper III:
\begin{eqnarray*}
\sigma_{FIT}&=&\frac{\sigma_{FITIN}\sigma_{FITOUT}}
{\sqrt{\sigma^2_{FITIN}+\sigma^2_{FITOUT}}}\\
\sigma_{CAL}&=&\sqrt{\frac{\sigma_{CALIN}^2+\sigma_{CALOUT}^2}{2}}\ \ \ \
\sigma=\sqrt{\sigma_{FIT}^2+\sigma_{CAL}^2}
\end{eqnarray*}


Given our complete open data policy anyone can try by himself to
compute other ages with our proper elements and family classification
data\footnote{http://hamilton.dm.unipi.it/astdys/index.php?pc=5}. However,
we recommend caution: ages computed with insufficient data could be
unreliable. It is also possible to improve ages (and decrease
uncertainty) by using our computed slopes but revising the Yarkovsky
calibration with a specific effort for each individual family.

\begin{figure}[h!]
\figfigincl{14 cm}{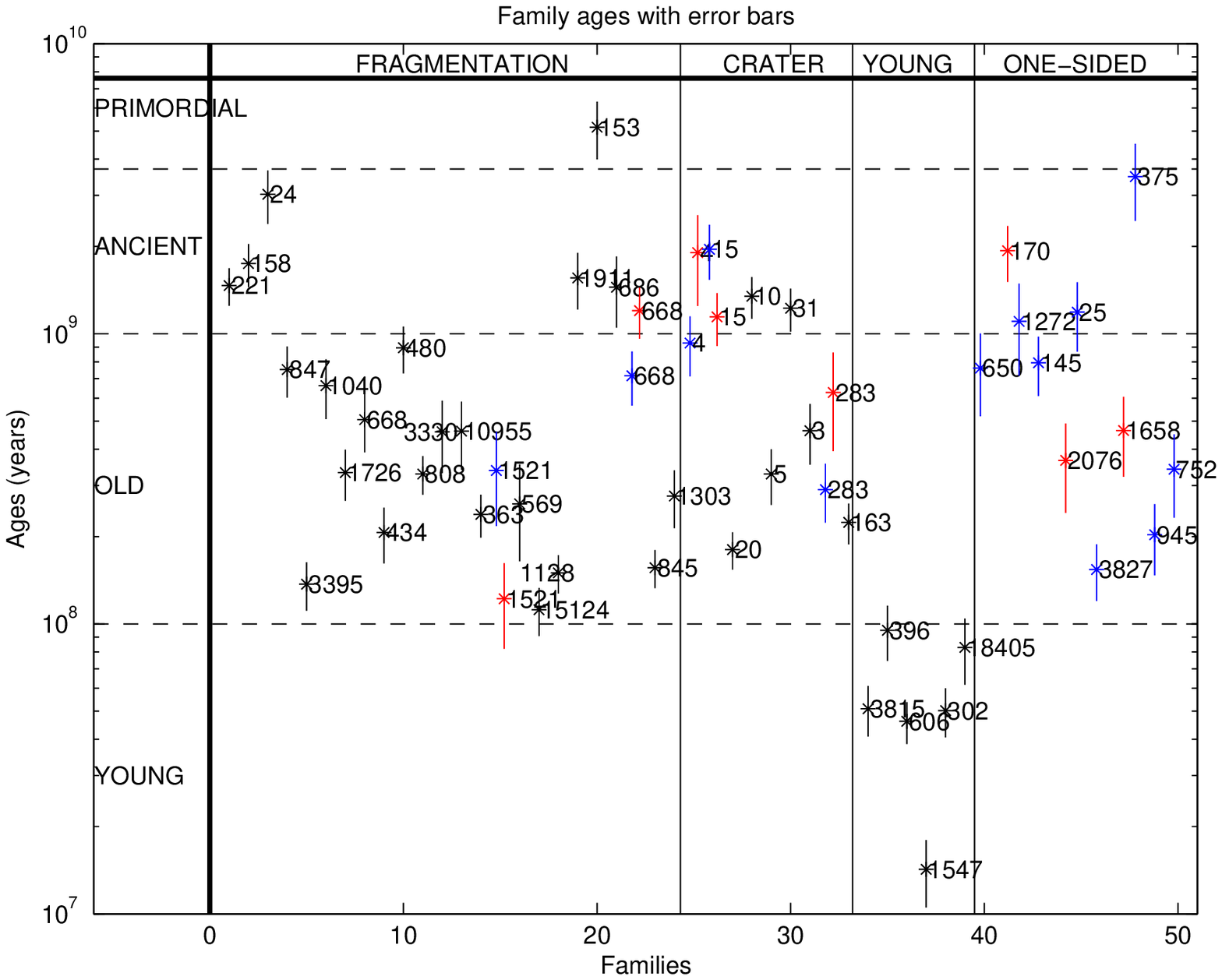}{Chronology of the asteroid families;
  the grouping on the horizontal axis correspond to fragmentation
  families, cratering families, young families and families with
  one-sided V-shape (be they cratering or fragmentation).}
\end{figure}

\section*{Acknowledgments}

The authors have been supported for this research by: ITN Marie Curie
``STARDUST -- the Asteroid and Space Debris Network''
(FP7-PEOPLE-2012-ITN, Project number 317185) (A.M., Z.K, B.N., and
G.T.), the Department of Mathematics of the University of Pisa (A.M.),
and the Ministry of Education, Science and Technological Development
of Serbia, under the project 176011 (Z.K., B.N.). F.S. is a fellow of
the CNES postdoctoral program - project called ``Asteroid observations
in the Gaia era''.


\section*{References}

\bibliographystyle{elsarticle-harv}

\section*{Appendix: Data for the ages estimation}

\begin{table}[h!]
 \centering
  \caption{Fit region: family number and name, explanation of the
    choice, minimum and maximum value of proper $e$, minimum and
    maximum value of the diameter selected for the inner and the outer
    side.}
  \label{tab:tablefite}
\medskip
  \begin{tabular}{|l|cr|cr|}
  \hline
            & \multicolumn{2}{|c|}{IN side} & \multicolumn{2}{c|}{OUT side}  \\
Number/name & Min proper $e$ & Min $D$ & Max proper $e$ & Min D \\
\hline
1911 Schubart & 0.16 &  5.00 & 0.22 & 6.67\\
153 Hilda     & 0.08 & 10.00 & 0.29 & 6.67\\
\hline
\end{tabular}
\end{table}

\begin{table}[h!]
 \centering
  \caption{Fit region: family number and name, explanation of the
    choice, minimum and maximum value of proper $a$, minimum and
    maximum value of the diameter selected for the inner and the outer
    side.}
  \label{tab:tablefita}
\medskip
  \begin{tabular}{|l|lll|lll|}
  \hline
            & \multicolumn{3}{c}{IN side}& \multicolumn{3}{|c|}{OUT side} \\
Number/name & Cause & Min $a$ & Min $D$& Cause & Max $a$ & Min D \\
\hline
110 Lydia      & 8/3J &  2.704  & 2.50 & Ceres   & 2.767 & 2.50\\
194 Prokne     & 3/1J &  2.52   & 3.33 &  FB     & 2.66  & 3.45\\
\hline
5 Astraea      & FB   &  2.555  & 2.00 & 2J+2S-1A & 2.61 & 2.22\\ 
283 Emma       & 9/4J &  3.03   & 6.67 & 11/5J   & 3.072 & 4.00\\
\hline
145 Adeona     & FB   &  2.58   & 3.33 & 8/3J    & 2.7   &      \\
25 Phocaea     & 7/2J &  2.26   & 2.50 & YORP?   & 2.45  &      \\
945 Barcelona  & FB   &  2.59   & 1.25 & 11/4J   & 2.65  &      \\
\hline
\end{tabular}
\end{table}

\begin{table}[h!]
 \centering
  \caption{Family albedos: number and name of the family, albedo of
    the largest body with its standard deviation and code for source
    (W=WISE, I=IRAS), maximum and minimum albedo values for computing mean,
    mean and standard deviation of the albedo of the members with
    $S/N>3$ WISE data. For the Hilda region we have used the values
    of the mean albedo and STD from \cite{grav2012}.}
  \label{tab:tablealbedo}
  \medskip
  \begin{tabular}{|l|ll|l|llll|}
    \hline
    & \multicolumn{2}{c|}{Largest Body}& &\multicolumn{4}{|c|}{Family Albedo} \\ 
    Number/name & Albedo & STD & Ref. & Min & Max & Mean & STD\\
    \hline

    110 Lydia     & 0.170 & 0.042 & W &      & 0.15 & 0.073 & 0.020\\
    1911 Schubart & 0.025 & 0.001 & I &      &      & 0.039 & 0.013\\
    153 Hilda     & 0.062 & 0.002 & I &      &      & 0.061 & 0.011\\
    194 Prokne    & 0.142 & 0.004 & W & 0.08 & 0.25 & 0.145 & 0.040\\
    \hline
    5 Astraea     & 0.245 & 0.051 & W & 0.10 & 0.50 & 0.269 & 0.076\\
    283 Emma      & 0.032 & 0.004 & W &      & 0.10 & 0.049 & 0.013\\
    \hline
    145 Adeona    & 0.043 & 0.013 & W &      & 0.10 & 0.062 & 0.010\\
    25 Phocaea    & 0.231 & 0.024 & I &      & 0.60 & 0.253 & 0.117\\
    945 Barcelona & 0.242 & 0.024 & I &      & 0.50 & 0.300 & 0.100\\
\hline
  \end{tabular}
\end{table}
\bigskip

\begin{table}[h!]
  \centering
  \caption{Slopes of the V-shape for the families in the 3/2 resonance: family number/name,
    side, slope $(S)$ in the ($e,1/D$) plane, inverse slope $(1/S)$,
    standard deviation of the inverse slope, ratio $OUT/IN$ of $1/S$,
    and standard deviation of the ratio.}
  \label{tab:tableslopese}
  \medskip
  \begin{tabular}{|l|lrrccc|}
  \hline
  Number/name  & Side & $S$    & $1/S$   & STD $1/S$  & Ratio & STD ratio \\
  \hline
  1911 Schubart & IN   &           -3.845 &           -0.260  & 0.028 &        &      \\
                & OUT  & \phantom{-}3.703 & \phantom{-}0.270  & 0.020 &  1.04  & 0.13 \\
  153 Hilda     & IN   &           -1.038 &           -0.963  & 0.084 &        &      \\
                & OUT  & \phantom{-}0.998 & \phantom{-}1.002  & 0.147 &  1.04  & 0.18 \\
  \hline
  \end{tabular}
\end{table}

\begin{table}[h!]
  \caption{Slopes of the V-shape for the families: family number/name,
    side, slope $(S)$ in the ($a,1/D$) plane, inverse slope $(1/S)$,
    standard deviation of the inverse slope, ratio $OUT/IN$ of $1/S$,
    and standard deviation of the ratio. Note the change of the family
    namesakes, 110 to 363, 194 to 686.}
  \label{tab:tableslopesa}
  \medskip
  \begin{tabular}{|l|lrrccc|}
  \hline
  Number/name  & Side & $S$              & $1/S$   & STD $1/S$  & Ratio & STD ratio \\
  \hline
 
  363 Padua    & IN   &           -5.972 &           -0.168  & 0.027 &        &      \\
               & OUT  & \phantom{-}7.844 & \phantom{-}0.128  & 0.112 &  0.76  & 0.14 \\
  686 Gersuind & IN   &           -1.758 &           -0.569  & 0.322 &        &      \\
               & OUT  & \phantom{-}1.874 & \phantom{-}0.534  & 0.138 &  0.94  & 0.58 \\
  \hline
  5 Astraea    & IN   &           -7.942 &           -0.126  & 0.029 &        &      \\
               & OUT  & \phantom{-}8.467 & \phantom{-}0.118  & 0.028 &  0.94  & 0.31 \\
  283 Emma     & IN   &           -6.046 &           -0.165  & 0.019 &        &      \\
               & OUT  & \phantom{-}2.814 & \phantom{-}0.355  & 0.112 &  2.15  & 0.72 \\
  \hline
  145 Adeona   & IN   &           -2.014 &           -0.497  & 0.058 &        &      \\
  25 Phocaea   & IN   &           -2.122 &           -0.471  & 0.047 &        &      \\
  945 Barcelona& IN   &          -13.445 &           -0.074  & 0.014 &        &      \\
  \hline
  \end{tabular}
\end{table}

\begin{table}[h!]
  \caption{Data for the Yarkovsky calibration: family number and name,
    proper semimajor axis $a$ and eccentricity $e$ for the inner and
    the outer side, $1-A$, density value $\rho$ at $D=1$ km, taxonomic
    type, a flag with values m (measured) a (assumed) g (guessed), and
    the relative standard deviation of the calibration.}
  \label{tab:tableyarkoparam}
  \medskip
  \begin{tabular}{|l|ll|ll|llccl|}
  \hline
  &\multicolumn{2}{|c|}{IN side}&\multicolumn{2}{c|}{OUT side}& & &  Tax&&\\
  Numb/name  & $a$& $e$& $a$& $e$& 1-A & $\rho$ & typ & Fl & STD\\
  \hline
  363 Padua    & 2.70 & 0.04 & 2.78 & 0.04 & 0.97 & 1.41  & C & m & 0.20 \\
  1911 Schubart& 3.963& 0.145& 3.967& 0.214& 0.98 & 1.41  & P & g & 0.30 \\
  153 Hilda    & 3.95 & 0.07 & 4.00 & 0.30 & 0.98 & 1.41  & P & g & 0.30 \\
  686 Gersuind & 2.52 & 0.175& 2.66 & 0.175& 0.95 & 2.275 & S & m & 0.20 \\
  \hline
  5 Astraea    & 2.56 & 0.18 & 2.60 & 0.19 & 0.93 & 2.275 & S & m & 0.20 \\
  283 Emma     & 3.03 & 0.115& 3.07 & 0.115& 0.98 & 1.41  & C & m & 0.20 \\
  \hline
  145 Adeona   & 2.58 & 0.162&      &      & 0.98 & 1.41  & C & m & 0.20 \\ 
  25 Phocaea   & 2.26 & 0.215&      &      & 0.92 & 2.275 & S & a & 0.25 \\
  945 Barcelona& 2.60 & 0.24 &      &      & 0.90 & 2.275 & S & m & 0.20 \\
  \hline
  \end{tabular}
\end{table}

\begin{table}[t]
  \caption{Age estimation for the families in the 3/2 resonance: family
    number and name, calibration $de/dt$, age estimation, uncertainty of the age
    due to the fit, uncertainty of the age due to the calibration, and
    total uncertainty of the age estimation.}
  \label{tab:tableages_dedt}
  \medskip
  \begin{tabular}{|l|ccrrrr|}
  \hline
  Numb/name  & Side & $de/dt$  & Age & STD$_{FIT}$ & STD$_{CAL}$ & STD$_{AGE}$ \\
             &          & $10^{-4}/My$ & My  & My       & My       & My       \\
  \hline
  1911 Schubart& IN  &           -1.68 &1547 & 165 & 464 & 492\\
               & OUT & \phantom{-}1.72 &1566 & 115 & 470 & 484\\
  153 Hilda    & IN  &           -1.83 &5265 & 461 &1580 &1645\\
               & OUT & \phantom{-}1.99 &5039 & 737 &1512 &1682\\
  \hline
  \end{tabular}
\end{table}

\begin{table}[h!]
  \caption{Age estimation for the families: family number and name,
    calibration $da/dt$, age estimation, uncertainty of the age due to
    the fit, uncertainty of the age due to the calibration, and total
    uncertainty of the age estimation.}
  \label{tab:tableages_dadt}
  \medskip
  \begin{tabular}{|l|ccrrrr|}
  \hline
  Numb/name  & Side & $da/dt$ & Age & STD$_{FIT}$ & STD$_{CAL}$ & STD$_{AGE}$ \\
             &            & $10^{-4} au/My$ & My  & My       & My       & My \\
  \hline 
  363 Padua    & IN  &           -5.90 & 284 &  46 &  57 &  73\\
               & OUT & \phantom{-}5.82 & 219 &  19 &  44 &  48\\         
  686 Gersuind & IN  &           -3.82 &1490 & 843 & 298 & 894\\
               & OUT & \phantom{-}3.62 &1436 & 371 & 287 & 469\\
  \hline
  5 Astraea    & IN  &           -3.72 & 339 &  79 &  68 & 104\\ 
               & OUT & \phantom{-}3.70 & 319 &  74 &  64 &  98\\ 
  283 Emma     & IN  &           -5.69 & 290 &  33 &  58 &  67\\
               & OUT & \phantom{-}5.66 & 628 & 197 & 126 & 234\\
  \hline
  145 Adeona   & IN  &           -6.25 & 794 &  92 & 159 & 184\\
  25 Phocaea   & IN  &           -3.97 &1187 & 117 & 297 & 319\\
  945 Barcelona& IN  &           -3.66 & 203 &  38 &  41 &  56\\
  \hline
  \end{tabular}
\end{table}

\end{document}